\documentclass[longauth]{aa}  

\usepackage{graphicx}
\usepackage{txfonts}
\usepackage{longtable}
\usepackage{lscape}
\usepackage{xcolor}
\usepackage{booktabs}
\usepackage{tabularx}
\usepackage{multirow}
\usepackage{amsmath}
\newcommand{\kms}{\,km\,s$^{-1}$} 
\newcommand{\ms}{\,m\,s$^{-1}$} 

\newcommand{\mearth}{$M_{\oplus}$}

\newcommand{\msun}{$M_{\odot}$}

\begin{document} 

   \title{The GAPS programme at TNG}

   \subtitle{LXXIX. New mass constraints for the infant planets in the V1298\,Tau system through a long-term radial velocity monitoring with HARPS-N}
   \author{M.~Damasso\inst{1}\thanks{\email{mario.damasso@inaf.it}} \and P.~Leonardi\inst{2,3} \and L.~Borsato\inst{3} \and S.~Benatti\inst{4} \and A.~Su{\'a}rez~Mascare{\~n}o\inst{5,6} \and A.~Bonfanti\inst{7} \and C.~Di Maio\inst{4} \and M.~Mallorqu\'{i}n\inst{8,5,6} \and L.~Fossati\inst{7} \and F.~Marzari\inst{2} \and D.~Nardiello\inst{2} \and S.~Desidera\inst{3} \and L.~Naponiello\inst{1} \and A.~Sozzetti\inst{1} \and G.~Andreuzzi\inst{9} \and M.~Basilicata\inst{10} \and A.~Bignamini\inst{11} \and A.~S.~Bonomo\inst{1} \and M.~Brogi\inst{12,1} \and P.~Giacobbe\inst{1} \and G.~Guilluy\inst{1} \and A.~F.~Lanza\inst{13} \and M.~Lodi\inst{9} \and L.~Malavolta\inst{2} \and L.~Mancini\inst{14,1} \and S.~Messina\inst{13} \and G.~Micela\inst{4} \and V.~Nascimbeni\inst{3} \and T.~Zingales\inst{2,3}
   }

   \institute{INAF -- Osservatorio Astrofisico di Torino, Via Osservatorio 20, I-10025 Pino Torinese, Italy
\and Dipartimento di Fisica e Astronomia, Università degli Studi di Padova, Vicolo dell’Osservatorio 3, I-35122 Padova, Italy
\and INAF -- Osservatorio Astronomico di Padova, Vicolo dell'Osservatorio 5, I-35122, Padova, Italy
\and INAF -- Osservatorio Astronomico di Palermo, Piazza del Parlamento 1, I-90134, Palermo, Italy
\and Instituto de Astrof\'{\i}sica de Canarias, c/ V\'ia L\'actea s/n, 38205 La Laguna, Tenerife, Spain
\and Departamento de Astrof\'{\i}sica, Universidad de La Laguna, 38206 La Laguna, Tenerife, Spain 
\and Space Research Institute, Austrian Academy of Sciences, Schmiedlstrasse 6, 8042 Graz, Austria
\and Light Bridges, S.L. Observatorio Astron\'omico del Teide, Carretera del Observatorio s/n, 38500 G\"u\'imar, Tenerife, Spain
\and Fundación Galileo Galilei-INAF, Rambla José Ana Fernández Pérez 7, 38712, Breña Baja, Spain
\and INAF -- Osservatorio Astronomico di Brera, Via E. Bianchi 46, 23807 -- Merate, Italy 
\and INAF -- Osservatorio Astronomico di Trieste, via Tiepolo 11, 34143 Trieste
\and Dipartimento di Fisica, Università di Torino, Via Pietro Giuria 1, 10125, Torino, Italy
\and INAF -- Osservatorio Astrofisico di Catania, Via S.~Sofia 78 - 95123 Catania, Italy
\and Dipartimento di Fisica, Università di Roma ``Tor Vergata'', Via della Ricerca Scientifica 1, 00133, Rome, Italy 
             }

   \date{Received 22 May 2026 / Accepted dd September 2026}

 
  \abstract
   {}
   {We aim to constrain the dynamical masses of the four confirmed transiting planets orbiting the infant ($20\pm10$ Myr) star V1298\,Tau (orbital periods $P_b\sim$24.1 d, $P_c\sim$8.2 d, $P_d\sim$12.4 d, and $P_e\sim$48.7 d), through a new analysis of a homogeneous radial velocity dataset. This system is a key benchmark for testing formation and early-stage evolutionary models of planets.}
   {We analysed more than 380 high-resolution spectra collected with the HARPS-N spectrograph over a time span of 4.5\,yr. To account for dominant stellar activity and transit-timing-derived planet-planet interactions, we analysed radial velocities extracted using two different approaches, and combined N-body simulations with multi-dimensional Gaussian process regression. We tested a set of spectroscopic and photometric activity diagnostics to mitigate the variability due to the stellar magnetic activity in the radial velocity time series.}
   {The mass determinations are sensitive to the choice of the radial velocity extraction method and activity diagnostics. For some specific models, we measured masses with a significance of 3.5--4$\sigma$ for planets b ($0.52^{+0.12}_{-0.14}$, $0.65^{+0.21}_{-0.18}$, $0.67\pm0.16$, and $0.78^{+0.19}_{-0.20}$ $M_{\rm Jup}$) and e (0.40$^{+0.14}_{-0.13}$, 0.66$^{+0.17}_{-0.18}$, and $0.71\pm0.20$ $M_{\rm Jup}$). These values are compatible with masses of giant planets, and are notably higher than those recently derived from transit timing variations alone, which indicate that these are low-density ``super-puffy'' planets. Injection-and-recovery simulations moderately support this conclusion for planet b. For planets c and d, we primarily provide updated mass upper limits.} 
   {Our analysis represents an exploratory study to address the complex issue of mass measurement in a multi-planetary system characterised by high levels of stellar activity. While firm conclusions cannot be drawn for V1298\,Tau, nonetheless in a few cases our results leave open the possibility that especially planet b is a massive giant rather than a sub-Neptune progenitor. This dichotomy may be solved by combining radial velocities and transit time variations in a more complex photo-dynamical modelling, with a larger sample of transits than currently available.}


  \titlerunning{Mass constraints for the planets in the V1298\,Tau system}
   \authorrunning{Damasso et al.}
   \maketitle
%

\section{Introduction}

Observational exoplanet science has advanced dramatically in the last few years, revealing a rich diversity of planetary systems in our galactic neighbourhood. A crucial frontier still remains largely unexplored: the detection and characterization of infant exoplanets, that we define as those newly formed and still evolving around stars less than a hundred million years old. These very young systems are the key to understanding how planets form and evolve at the earliest stages, yet they remain under-represented in the current general exoplanet population. Particularly valuable are planets orbiting close to their host star that can be detected via transit photometry, because in principle their bulk density may be measured through a spectroscopic follow-up, constraining their physical structure and orbital architecture. However, close-in infant planets represent less than 0.5\% of the known exoplanet population \footnote{\url{https://exoplanetarchive.ipac.caltech.edu/}}. Although recent studies have begun exploring this area \citep[e.g.][]{vach_2024AJ....167..210V, Dai_2024AJ....168..239D,fernandes_2025AJ....169..208F}, it is still too early for a systematic demographic comparison between infant and mature systems.
Therefore, a thorough characterisation of each individual infant system is expected to bring a high reward, although it requires a huge investment in terms of observational and computing resources.         
In this context, infant multi-planet systems are particularly valuable: they allow us to study the physical and orbital diversity of planets forming and rapidly evolving within the same environment. Among the few of such systems known to date, the one around the $20\pm10$ Myr-old K1 star V1298\,Tau (also known as BD+19 656 and K2-309, $V$=10.12$\pm$0.05, $d$=109.5$\pm$0.7 pc; \citealt{Suarez2022NatAs...6..232S}) undoubtedly represents a benchmark target to test models of planetary formation and evolution. V1298\,Tau hosts four transiting planets discovered by \textit{Kepler/K2} \citep{david2019AJ....158...79D,David2019ApJ...885L..12D}, with a not yet confirmed fifth planet candidate possibly detected by CHEOPS \citep{2023A&A...680A...8D}. 

In Table \ref{tab:noteliteraturesummary} we summarise the studies published so far that were focused on the determinations of the ephemeris, radius, and mass of the four confirmed planets in the V1298\,Tau system. Datasets and analysis methods used in each study are also reported. In Table \ref{Table:Summary_planets} we report a selection of the published results, in order to appreciate the various efforts that have been made over the years, and track how the results have changed over time. Key questions surrounding the system since its 2019 discovery include determining the orbital period of outermost planet e, measuring planetary masses and radii, or detecting and characterizing their atmospheres. The recent work by \cite{livingston2026Natur.649..310L} (hereafter L26) stands out because it provides answers to the first two questions. They succeeded in pinning down the orbital period of V1298\,Tau\,e, revealing that the outermost planets b and e are close to the 2:1 orbital resonance, and in measuring the dynamical masses through the modelling of the transit time variations (TTVs) detected for all four confirmed planets. L26 derived masses in the super-Earth to sub-Neptune regime. Combined with the large planet radii, these results indicate very low densities, suggesting that photo-evaporation will eventually shrink these planets to the sizes of mature super-Earths and sub-Neptunes. These findings have important implications for the formation and early-stage evolutionary channels of the four planets, such as that the innermost planets c and d fit the boil-off scenario, a phase of extreme atmospheric loss occurring after the protoplanetary disk dispersal, and that planets b and e are not rapidly evolved giant-like planets as found by \cite{Suarez2022NatAs...6..232S} through the analysis of radial velocities (RV), a result later questioned by \citealt{Blunt2023AJ....166...62B}. Therefore, it is crucial to confirm the results of L26 with more data and new analysis methods.    

It is well-known that a major obstacle to the characterisation of close-in infant planets, such those orbiting V1298\,Tau, is the high level of stellar magnetic activity, which hampers both photometry and RVs. Host stars exhibit strong variability driven by activity produces Doppler signals in RVs with amplitudes up to two orders of magnitude larger than planetary signals. Any attempt to mitigate these dominant jitter sources requires dedicated and optimised observational strategies, and advanced modelling techniques, including methods for measuring the RVs from high resolution spectra, as it has been explored in other studies involving young systems \citep[e.g. ][]{barragan2023MNRAS.522.3458B,barragan2026MNRAS.546ag087B,Damasso2023A&A...672A.126D,Damasso2024A&A...690A.235D}.

In this work, we present results from modelling more than 380 RVs measured from high-resolution spectra collected by the HARPS-N spectrograph over a timespan of nearly 1700 days, with the goal of providing new constraints to the dynamical masses of the four transiting planets. This represents the largest and homogeneous spectroscopic dataset of V1298\,Tau that has been collected so far. To pursue our goal, we use complex analysis and fitting techniques to reduce the dominant contribution of stellar magnetic activity, simultaneously modelling planet-planet gravitational perturbations via $N$-body integration, laying the foundations for a follow-up investigation through a photo-dynamical modelling. The paper is organised as follows. In Sect. \ref{sec:dataset} we describe the spectroscopic and photometric data used in this study. In Sect. \ref{sec:modellingframework} we present the adopted working hypothesis and methods. The results of the RV modelling are discussed in Sect. \ref{sec:modellingresults}, and conclusions are drawn in Sect. \ref{sec:conclusions}.

\begin{table*}[!htbp]
    \centering
    \scriptsize
    \renewcommand{\arraystretch}{1.3} 
     \caption{Summary of datasets, modelling approaches, and relevant literature notes for the V1298\,Tau planetary system.}
    \begin{tabularx}{\textwidth}{l p{2.8cm} p{2.8cm} X}
        \hline\hline
        Reference & Datasets Used & Models and Approach & Key Notes and Findings \\ 
        \midrule
        \cite{David2019ApJ...885L..12D} & Photometry: $K2$ & Discovery paper; transit analysis & Initial detection of planets b, c, d, and e in the V1298\,Tau system. Only one transit of e was observed, and the orbital period $P_{\rm e}$ could not be constrained. \\
        \midrule
        \cite{Suarez2022NatAs...6..232S} & Photometry: $K2$ and LCOGT; RVs: HARPS-N, CARMENES, HERMES and SES & Simultaneous transit and RVs fitting. GP PQP2 kernel for stellar activity mitigation. & Detection of two periodic Keplerian signals compatible with the orbits of planets b and e, deriving 0.64 $M_{\rm Jup}$ and 1.16 $M_{\rm Jup}$, respectively. \\
        \midrule
        \cite{Feinstein2022ApJ...925L...2F} & Photometry: $K2$ and TESS & Simultaneous transit + activity modelling & Calculated a grid of discrete values for $P_{\rm e}$, each compatible with the observed $K2$ and TESS transits. \\
        \midrule
        \cite{Sikora_2023} & Photometry: $K2$, TESS, LCOGT; RVs: Archival + MAROON-X & Joint transit+RV modelling; GP kernels (SHO and QP) &
        2.6$\sigma$ significant RV detection of planet e with a mass 0.66 $M_{\rm Jup}$, assuming an orbital period of $\sim$ 46 days.   \\
        \midrule
        \cite{Finociety2023MNRAS.526.4627F} & RVs: SPIRou nIR spectra & GP QP kernel for activity modelling & Detection of the RV signature of planet e, with an orbital period of $\sim$ 53 days, and mass 1 $M_{\rm Jup}$. \\ 
        \midrule
        Damasso et al. (2023b) & Photometry: CHEOPS  & Transit follow-up & Transit-like signal observed with CHEOPS, potentially due to a fifth planet-size companion. \\
        \midrule
        \cite{Barat2024NatAs.tmp...86B} & HST/WFC3 transmission spectra  of V1298\,Tau\,b & Atmospheric modelling & Derived mass $m_b = 24\,M_\oplus$. This value represents an upper limit if the atmosphere is partly cloudy or hazy. \\
        \midrule
        \cite{Barat_2025} & JWST/NIRSpec G395H transmission spectra of V1298\,Tau\,b & Atmospheric modelling & By combining data from JWST and HST, derived mass $m_b$=12$\pm$1 and 15$\pm$1.7 $M_\oplus$, using free retrieval and grid modelling respectively.  \\
        \midrule
        \cite{livingston2026Natur.649..310L} & Photometry: $K2$, TESS, Spitzer, MuSCAT and LCO & TTV modelling & Planetary masses for all four transiting planets were derived. The orbital period $P_{\rm e}$ was pinned down. \\
        \bottomrule
    \end{tabularx}
    \label{tab:noteliteraturesummary}
\end{table*}

\section{Description of the data} \label{sec:dataset}

\subsection{HARPS-N spectroscopic follow-up} \label{sec:harpndata}
Our initial spectroscopic dataset consists of 394 spectra collected with the HARPS-N spectrograph \citep{Cosentino2012} from 1 Mar 2019 to 27 Sept 2023. Spectra collected up to April 2020 were analysed in \cite{Suarez2022NatAs...6..232S}, and in this work we increase the dataset with 262 additional observations that were acquired as part of the program aimed at characterization of planets around young stars in the framework of the GAPS (Global Architecture of Planetary Systems) collaboration \citep[e.g. ][]{carleo2020A&A...638A...5C}. The total sample is characterised by a median signal-to-noise ratio S/N=56.6, measured at a reference wavelength of 5500\,\AA\, (echelle order 46). Initially, we discarded three spectra corresponding to epochs 2458888.457 BJD$_{\rm TDB}$ (8 Feb 2020) because of bad weather, and 2459575.449 BJD$_{\rm TDB}$ (26 Dec 2021), 2459831.730 BJD$_{\rm TDB}$ (8 Sept 2022) because of a significant mismatch with the expected systemic velocity.  
We have taken into account the inherent complexities in determining RVs for young, active, and rapidly rotating stars, using two different methods, based on the cross-correlation function (CCF) and template matching. The dependence of the RV modelling results from the RV extraction recipes has been heuristically investigated in other GAPS works related to young systems \citep[e.g.][]{Damasso2020A&A...642A.133D,Damasso2023A&A...672A.126D,Damasso2024A&A...690A.235D}. The methods were applied to spectra first processed with the HARPS-N Data Reduction Software (\texttt{DRS}; version 3.7) through the \texttt{YABI} workflow interface \citep{YABI}\footnote{\url{ https://ia2.inaf.it}} maintained by the Italian center for Astronomical Archive (IA2). The \texttt{DRS} provides automatically RVs and activity diagnostics, such as the bisector inverse slope (BIS), which is derived from the CCF calculated by adopting a reference mask for a star with spectral type K5 and a half-window width of 100 \kms, which is not the default value adopted in the \texttt{DRS} pipeline. \citep{baranne_1996A&AS..119..373B,bouchy_2001A&A...374..733B}. However, this method is not optimal for fast-rotating, highly active stars like V1298\,Tau, as its broad CCFs deviate significantly from a Gaussian profile. We therefore did not use the RVs calculated by the \texttt{DRS} (following \citealt{Suarez2022NatAs...6..232S}); nonetheless, we retained the BIS as an activity diagnostic to filter stellar variability. 
We used the publicly available template-matching algorithm \texttt{serval} developed in \texttt{python} \citep[version 2022-01-26;][]{serval2018A&A...609A..12Z} using the following keywords: \texttt{-safemode 2 -niter 4 -snmin 25 -ofac 1.00 -vrange 30}. Of the initial 391 spectra, we discarded five due to low signal-to-noise ratio $(S/N < 25)$ and three more because their corresponding BIS values had been removed. Although correctly processed by \texttt{serval}, three spectra (at epochs 2459829.767, 2459989.330, and 2460173.700 BJD$_{\rm TDB}$) were excluded to maintain a strict 1:1 epoch correspondence between RVs and the BIS activity diagnostic (Sect. \ref{sec:modellingframework}). The final set consists of 383 good spectra that cover a timespan of 1671 days, and the RVs have a dispersion and median internal error $\sigma_{\rm RV}$ of 236 and 9 \ms, respectively. One of the products of \texttt{serval} is the chromatic index (CRX, in units of velocity per natural logarithm of the wavelength ratio, the latter expressed in units of the Neper number e, indicated as $Np$). We used the CRX as an activity diagnostic \citep[see e.g. ][]{serval2018A&A...609A..12Z} for the analysis described in Sect. \ref{sec:modellingframework}. 

Finally, we extracted the RV time series following the method described by \citet{simola2019}. In short, rather than using a Gaussian function to model the CCFs, these are fitted using skew normal (SN) functions, which contain not only a location and a scale parameter (i.e. the counterparts of the mean and standard deviation of a Normal function), but also a skewness parameter ($\gamma_{\rm SN}$) that quantifies the CCF asymmetry. 
From the initial 391 spectra, similarly to \texttt{serval} five spectra with a S/N ratio lower than 25 were discarded, and two more, observed at 2459829.767 and 2459990.457 BJD$_{\rm TDB}$, because the SN-fit routine did not converge. We ended up with an RV time series made of 384 data points, from which we removed two more spectra corresponding to outliers identified in the BIS dataset. In conclusion, the final dataset consists of 382 RVs, one data point less than the \texttt{serval} dataset, corresponding to the spectrum taken at 2459990.457 BJD$_{\rm TDB}$. The RVs have a dispersion and median internal error $\sigma_{\rm RV}$ of 289 and 17 \ms, respectively, both higher than the \texttt{serval} dataset. Hereafter, we will refer to this as the \texttt{snfit} dataset. 

\subsubsection{Frequency spectrum analysis}
The RVs extracted with \texttt{serval} and \texttt{snfit} were analysed in a similar way to investigate any dependence of the results from the RV extraction methods.
We show the RV time series and their generalised Lomb-Scargle periodograms \citep[GLS;][]{2009A&A...496..577Z} in the first two rows of Fig. \ref{fig:rvacttimeseriesgls}. The frequency content analysis reveals very significant peaks at the stellar rotation period ($P_{\rm rot}$$\simeq$2.9 d) and its first harmonic, as expected \citep{Suarez2022NatAs...6..232S}.
In Fig. \ref{fig:rvacttimeseriesgls} we also show the time series and corresponding GLS periodograms of the activity diagnostics considered in our study. In addition to the BIS and CRX index, we also analysed the chromospheric activity indicators derived from the H$_\alpha$ and \ion{Ca}{ii} H$\&$K spectral lines using the code \texttt{actin2} \citep{2018JOSS....3..667G,2021A&A...646A..77G}. The S-index calculated from the \ion{Ca}{ii} H\&K line doublet \citep{vaughan1978} was calibrated to the Mount Wilson scale \citep[see Sect. 4.2 in ][]{2021A&A...646A..77G} and is indicated as $S_{\rm MW}$. Their periodograms show significant peaks at the stellar rotation period, as seen for the RV time series, therefore these diagnostics appear as an effective choice to mitigate the activity term in a joint fit with the RVs. The H$\alpha$ and $S_{\rm MW}$ indexes also show evidence for a long-term modulation. 

\subsection{ASAS-SN photometry} \label{sec:asas}
\begin{figure*}
    \centering
    \includegraphics[width=\textwidth]{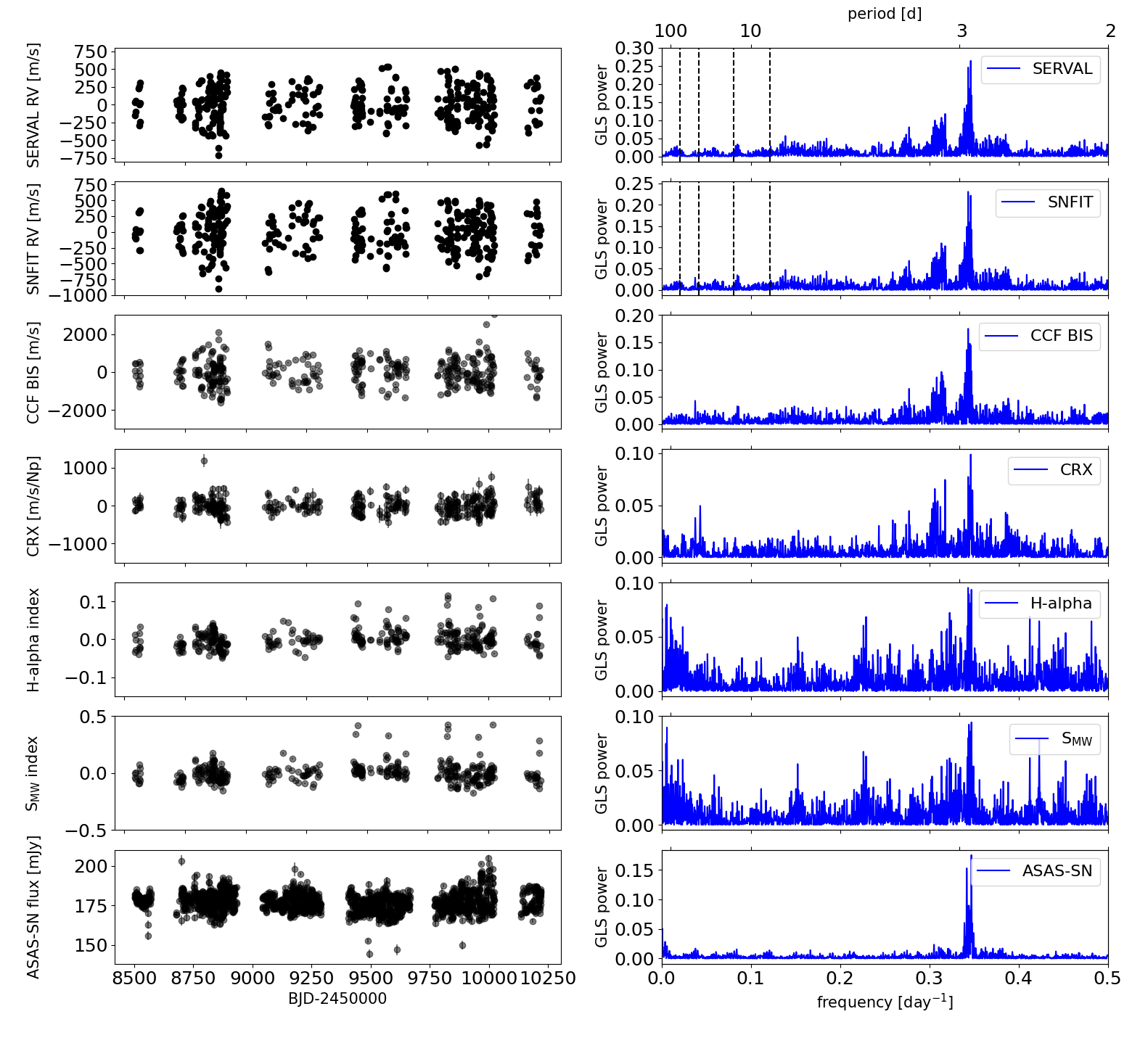}
    \caption{Radial velocity and spectroscopic activity diagnostics time series of V1298\,Tau (left panels), and corresponding GLS periodograms (right panels), with the vertical dashed lines in the RV panels indicating the orbital periods of the four transiting planets. }
    \label{fig:rvacttimeseriesgls}
\end{figure*}

We analysed the optical light curve (\textit{g}-band) of V1298\,Tau collected by the All-Sky Automated Survey for Supernovae (ASAS-SN) survey \citep{2014ApJ...788...48S,2023arXiv230403791H} during the time span covered by the HARPS-N follow-up. The data were downloaded from the SkyPatrol database\footnote{\url{https://asas-sn.osu.edu/}}. We discarded bad data indicated with a magnitude of 99.99 in the original file. The time series and GLS periodogram are displayed in the last panel of Fig. \ref{fig:rvacttimeseriesgls}. Very significant peaks are detected at the stellar rotation period and its first harmonic, and the periodogram shows the lowest noise level compared to other activity indicators. The ASAS-SN light curve is used in our modelling framework as a tracer of stellar activity (Sect. \ref{sec:modellingframework}), similarly to the spectroscopic activity diagnostic.  

\section{Working hypothesis and methods} \label{sec:modellingframework}
Our work is based on the long-term spectroscopic dataset collected only with HARPS-N. Other datasets collected with the spectrographs CARMENES, STELLA, HERMES \citep{Suarez2022NatAs...6..232S}, MAROON-X \citep{Sikora_2023}, and SPIRou \citep{Finociety2023MNRAS.526.4627F}, are not included in this study because they have a significantly lower number of RVs, either a lower precision, a shorter timespan overlapping with that covered by HARPS-N, and a less dense sampling than HARPS-N data. Therefore, we considered the use of the homogeneous and large HARPS-N dataset as the best option to investigate this challenging target, without increasing the number of free parameters in the models (such as offset and uncorrelated jitter for each instrument).

Our analysis is based on a number of hypotheses. We assume that the V1298\,Tau system is composed by the four planets observed by \textit{Kepler/K2} and TESS, and we do not consider the presence of a fifth companion, such as the transiting candidate identified with CHEOPS \citep{2023A&A...680A...8D}, because its existence is currently not supported by other data. This observed transit-like event occurred approximately 11 days later than the transit time predicted for planet e from our dynamical model (Leonardi et al. in prep.), making a direct association between signal and planet very unlikely. Given the detection of TTVs for the four planets due to their near-resonant state (L26), we used the \texttt{python} interface of the publicly available $N$-body integrator \texttt{trades}\footnote{\url{https://github.com/lucaborsato/trades}} \citep[][]{trades2014A&A...571A..38B,trades2019MNRAS.484.3233B,trades2024A&A...689A..52B}, that takes planet-planet gravitational interactions into account when modelling the system's dynamics, to fit the RV time series.  The free parameters used for the \texttt{trades} module are the orbital period $P$, the eccentricity e, the argument of pericenter $\omega$, the mean longitude $\lambda$, the inclination angle of the orbital plane $i$, and the mass $m$ (in units of stellar mass), for each of the four planets. The longitude of the ascending node $\Omega$ was fixed to 180 deg for each planet. The mean longitude is used to derive the actual parameter passed to \texttt{trades}, that is the mean anomaly $M=\lambda-\omega-\Omega$.
For the planetary orbital periods $P$ we used Gaussian distributions with the mean and variance corresponding to the best-fit results of L26, except for V1298\,Tau\,b, for which we used a revised value obtained from fitting part of our own larger TTV dataset that includes CHEOPS observations not yet public (Leonardi et al. in prep.). For the inclination angles of the orbital planes we used priors based on our reanalysis of the published photometric transits, and those from unpublished CHEOPS data of Leonardi et al. (in prep.). The epoch of the first RV measurement (2458544.398 BJD$_{\rm TDB}$) was used as the start time of the orbital integration and reference epoch for the values of the orbital parameters. Each simulation of the $N$-body integration covers the timespan of the HARPS-N data, and the time step used to evaluate the output parameters (and checking for transits, although here we do not use this information) is set to 1/10 of the orbital period of the innermost planet c. 
The mass of the host star was fixed to 1.17 \msun\, \citep{Suarez2022NatAs...6..232S}.  

To filter out the largely dominant stellar activity contribution to the RV variability, we tested different models based on Gaussian process (GP) regression. We used both one-dimensional \citep[e.g.][]{haywood2014MNRAS.443.2517H} and multi-dimensional \citep[e.g.][]{rajpaul15} GP approaches, by testing different datasets as tracers for the stellar activity. For all the tested GP models, we adopted the quasi-periodic (QP) covariance matrix with the generic element defined as  
\begin{gather} 
\label{eq:eqgpqpkernel}
k_{QP}(t, t^{\prime}) = h^2\exp\Bigg[-\frac{(t-t^{\prime})^2}{2\tau^2} - \frac{\sin^{2}\Bigg(\pi(t-t^{\prime})/\theta\Bigg)}{2w^2}\Bigg] +  \\
+\, \left(\sigma^{2}_{\rm RV}(t)+\sigma^{2}_{\rm jit}\right)\delta_{t, t^{\prime}}\; \nonumber
\end{gather}
Here, $t$ and $t^{\prime}$ denote two different epochs of observations, $\sigma_{\rm RV}$ is the RV uncertainty, and $\delta_{t, t^{\prime}}$ is the Kronecker delta. Uncorrelated noise, either instrumental and/or astrophysical, is taken into account by adding a constant jitter term $\sigma_{\rm jit}$ in quadrature to $\sigma_{\rm RV}$. The GP hyper-parameters are $h$, which denotes the scale amplitude of the correlated signal;  $\theta$, which represents the periodic time scale of the correlated signal and corresponds to the stellar rotation period, which is well constrained in the case of V1298\,Tau \citep[e.g. 2.91$\pm$0.05 d; ][]{Suarez2022NatAs...6..232S}; $w$, which describes the ``weight'' of the rotation period harmonic content within a complete stellar rotation (i.e. a low value of $w$ indicates that the periodic variations contain a significant contribution from the harmonics of the rotation period); and $\tau$, which represents the decay timescale of the correlations. 

We used the publicly available \texttt{python} module \texttt{george} v0.2.1 \citep{Ambikasaran2015} to perform the GP 1D regression. For the GP multi-dimensional approach, we used the code \texttt{pyaneti} \citep{pyaneti,pyaneti2_2022MNRAS.509..866B}. The theoretical formalism that we adopted for the multi-dimensional GP approach is the same as described in the works by \cite{rajpaul15} and \cite{pyaneti2_2022MNRAS.509..866B}, that we briefly summarise.
Given a set of N time-series $S_{\rm 1,...N}$ (with N$\geq$2), each one with M points taken at the epochs $t_{\rm 1,...M}$, the framework uses the same function $G(t)$ and its temporal derivative $\dot{G}(t)$ to model the activity-induced signal in each time series $S$. $G(t)$ is a latent unobserved variable which represents the activity term, and is described by a zero-mean GP with covariance term given by Eq. (\ref{eq:eqgpqpkernel}), with the hyper-parameter $h$ fixed to 1 in that case. Each time-series can be described by the set of $N$ equations like
\begin{gather}
\label{eq:multigp}
S_i = A_iG(t) + B_i\dot{G}(t)
\end{gather}
where $i$=1...$N$, and A$_i$, B$_i$ are free parameters. In our case, we explored the 2- and 3-dimensional cases ($N$=2,3), where one dataset is represented by the RVs, and the others by selected activity diagnostics. The multi-dimensional GP framework has been efficiently used to mitigate high stellar activity contributions to the RV variability in dataset of young planetary systems \citep[e.g.][]{barragan2023MNRAS.522.3458B,barragan2026MNRAS.546ag087B}.  
We explored the full (hyper-)parameter space using the publicly available Monte Carlo (MC) nested sampler and Bayesian inference tool \texttt{MultiNest v3.10} (e.g., \citealt{Feroz2019}), through the \texttt{pyMultiNest} wrapper \citep{Buchner2014}. The modules \texttt{trades}, \texttt{george}, and \texttt{pyaneti} were integrated within the MC nested sampler framework. Orbital solutions that result in planetary close encounters during the timespan covered by the simulations are excluded by imposing a likelihood function equal to -$\infty$.

To select the possibly most promising set of activity indicators to be used jointly with the RVs in the multi-dimensional GP regression, first we verified if a significant correlation exists between the RV time series and that of the BIS, CRX, H$_\alpha$, and S$_{\rm MW}$ spectroscopic indexes. The results for the \texttt{serval} RVs are shown in Fig. \ref{fig:rv_vs_act_indicators}. There is a strong anti-correlation with the BIS and CRX activity indicators ($\rho_{\rm Pearson}$= $-$0.84 and $-$0.68, respectively), while the RVs do not show correlation with the H$_\alpha$ and S$_{\rm MW}$ indexes. This could be explained by an RV variability produced by spectral line distortions due dark spots and bright faculae, rather that the quenching of convective blueshifts as it happens in low-activity stars such as the Sun. We obtained similar results for the \texttt{snfit} dataset, without considering the CRX index in this case because this is derived from the RVs calculated by the \texttt{serval} pipeline. Then, we modelled the time series of the spectroscopic activity indicators with a MC GP QP regression, also including the ASAS-SN light curve. We compared the best-fit results for the hyper-parameters with those for the RVs (\texttt{serval} and \texttt{snfit}) obtained by applying the same model (Table \ref{tab:gpactivityindicator}), in order to verify which activity diagnostics show time variability with similar properties to that of the RV time series. We found that the BIS, CRX, and ASAS-SN data show a QP pattern characterised by GP hyper-parameters $\theta$, $w$, and $\tau$ in good agreement with those of the RVs, while the rotation period $\theta$ is not well constrained for the S$_{\rm MW}$ index (although the periodogram shows a peak at the rotation period), and the timescale $\tau$ is much shorter for the H$\alpha$ index. The results for the S$_{\rm MW}$ and H$\alpha$ index could be influenced by the presence of low-frequency signals, as shown by their GLS periodograms (Fig. \ref{fig:rvacttimeseriesgls}).

Based on the outcome of this analysis, we selected BIS, CRX, H$\alpha$, and ASAS-SN photometry as the activity diagnostics to be tested in a multi-dimensional GP approach. We interpolated the ASAS-SN photometry to the epochs of the RV measurements to make it usable with \texttt{pyaneti}. This interpolation was necessary because the available version of \texttt{pyaneti} when we have carried out the analysis required RVs and activity indicators to share identical observational epochs. This limitation has been overcome since the work of \citealt{barragan2026MNRAS.546ag087B}. The photometric flux at the epochs of the HARPS-N spectra was estimated by using the best-fit GP QP solution (see Table \ref{tab:gpactivityindicator} and Fig. \ref{fig:asasinterp}). Given the dense sampling of the light curve, the GP regression allows for a robust predictive capability at the epochs of the RV observations. The predicted ASAS-SN light curve, and its GLS periodogram, are shown in Fig. \ref{fig:glsasas2}.

We summarise in Table \ref{tab:priors} the prior distributions adopted for the free parameters in our GP-based models. The coefficient $B_{\rm ASAS-SN}$ of the derivative term in Eq. (\ref{eq:multigp}) for the ASAS-SN photometry was fixed to zero \citep[see ][]{rajpaul15}.

\begin{table*}
\centering
\tiny
\caption{Prior distributions for the free parameters of our models.}
\label{tab:priors}
\begin{tabular}{l l l}
\hline\hline
 & Planetary parameter & Prior distribution$^{(a)}$ \\ 
\midrule
\textit{Planetary Parameters} & Period $P$ [d] & planet b: $\mathcal{N}(24.14049, 0.00021)$; planet c: $\mathcal{N}(8.24981, 0.00039)$ \\
(\texttt{trades} $N$-body)$^{(b)}$ & & planet d: $\mathcal{N}(12.40103, 0.00066)$; planet e: $\mathcal{N}(48.68028, 0.00021)$ \\
\addlinespace[2pt]
& Eccentricity e & $\mathcal{N}(0, 0.1)^{(c)}$ (all planets) \\
& Inclination $i$ [deg] & planet b: $\mathcal{N}$(88.5,0.9); planet c; $\mathcal{N}$(89.0,0.7); planet d: $\mathcal{N}$(89.0,0.5); planet e: $\mathcal{N}$(89.0,0.5) \\
& Argument of pericenter $\omega$ [deg] & $\mathcal{U}(0, 360)$ (all planets) \\
& Mean longitude $\lambda$ [deg] & $\mathcal{U}(0, 360)$ (all planets)$^{(d)}$  \\
& Long. of the ascending node $\Omega$ [deg] & 180 (fixed, all planets) \\
& Mass ratio $m/M_\star$ & $\mathcal{U}(0, +0.01)$ (all planets) \\
\midrule
\textit{GP QP} & $\theta$ [d] & $\mathcal{U}(2.7, 3.2)$ \\
(Activity Model) & $\tau$ [d] & $\mathcal{LU}(0, 10)$ \\
& $w$ & $\mathcal{U}(0, 1)$ \\
& $h$ [\ms] (1D, only RV)  & $\mathcal{U}(0, 500)$ \\
& $A_{\rm RV}, B_{\rm RV}$ [\ms] (Multi-dim) & $\mathcal{U}(-500, 500)$ \\
& $A_{\rm BIS}, B_{\rm BIS}$ [\ms] (Multi-dim) & $\mathcal{U}(-500, 500)$ \\
& $A_{\rm CRX}, B_{\rm CRX}$ [m/s/Np] (Multi-dim) & $\mathcal{U}(-500, 500)$ \\
& $A_{H\alpha}, B_{H\alpha}$ & $\mathcal{U}(-0.5, 0.5)$ \\
& $A_{\rm ASAS-SN}^{(e)}$ [mJy] & $\mathcal{U}(-30,30)$  \\
\midrule
\textit{Instrumental parameters} &  offset, $\gamma$, [RV and BIS: \ms; & RV: $\mathcal{U}(-500, 500)$; BIS: $\mathcal{U}(-300,300)$; CRX: $\mathcal{U}(-50,50)$; ASAS-SN: $\mathcal{U}(-30,30)$;\\
& CRX: m/s/Np; ASAS-SN: mJy] &  $H\alpha$: $\mathcal{U}(-1,1)$ \\
\addlinespace[2pt]
& uncorrelated jitter, $\sigma$, [RV and BIS: \ms; & RV: $\mathcal{U}(0, 300)$; BIS: $\mathcal{U}(0, 1000)$; CRX: $\mathcal{U}(0, 500)$, ASAS-SN: $\mathcal{U}(0, 10)$; $H\alpha$: $\mathcal{U}(0, 1)$  \\
& CRX: m/s/Np; ASAS-SN: mJy] \\
\bottomrule
\end{tabular}
  \tablefoot{
     \tablefoottext{a}{$\mathcal{N}$: Gaussian prior; $\mathcal{U}$: Uniform prior; $\mathcal{LU}$: Uniform prior in the natural logarithm space.}\\
     \tablefoottext{b}{These parameters refer to the reference epoch corresponding to the first RV measurement.}
    \tablefoottext{c}{This corresponds to an half-Gaussian prior (only values $e>0$ are sampled), which was adopted based on the results of \cite{vaneylen2019AJ....157...61V}. Very low eccentricities were determined by L26.}\\
    \tablefoottext{d}{The mean longitude $\lambda$ is used with $\omega$ and $\Omega$ to calculate the mean anomaly $M=\lambda-\omega-\Omega$, which is the input parameter given to \texttt{trades}.}\\
    \tablefoottext{e}{The coefficient of the derivative term $B_{\rm ASAS-SN}$ was fixed to 0.}}    
\end{table*}

\section{Results and discussion}  \label{sec:modellingresults}

\subsection{Planetary mass measurements}
\begin{figure*}
    \centering
    \includegraphics[width=0.95\textwidth]{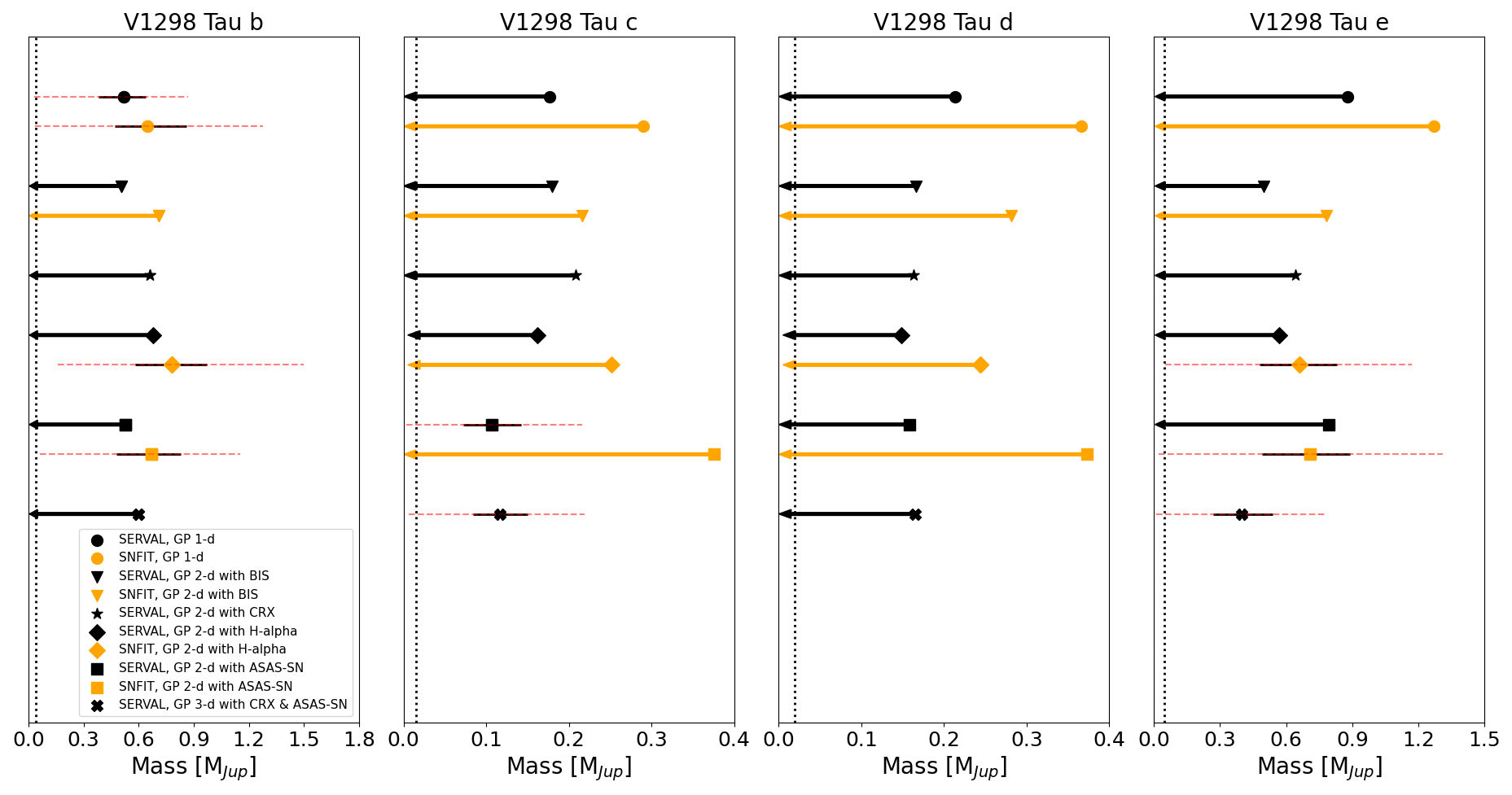}
    \caption{Dynamical masses of the planets in the V1298\,Tau system for each test model summarised in Table \ref{tab:massresults}. The ranges 15.87$^{\rm th}$--84.14$^{\rm th}$ (1$\sigma$) and 0.135$^{\rm th}$--99.865$^{\rm th}$ (3$\sigma$) percentiles of the marginalised posteriors are indicated around the value corresponding to the 50$^{\rm th}$ percentile by solid black and dashed red lines, respectively, for the masses determined with a significance better than three times the 1$\sigma$ lower error. For all the other mass values, the upper limits (99.865$^{\rm th}$ percentile) are indicated with a left-pointing arrow. Vertical dotted vertical lines indicate the masses measured by L26.}
    \label{fig:summarymasses}
\end{figure*}

\begin{figure*}[h]
    \centering
    \includegraphics[width=0.95\textwidth]{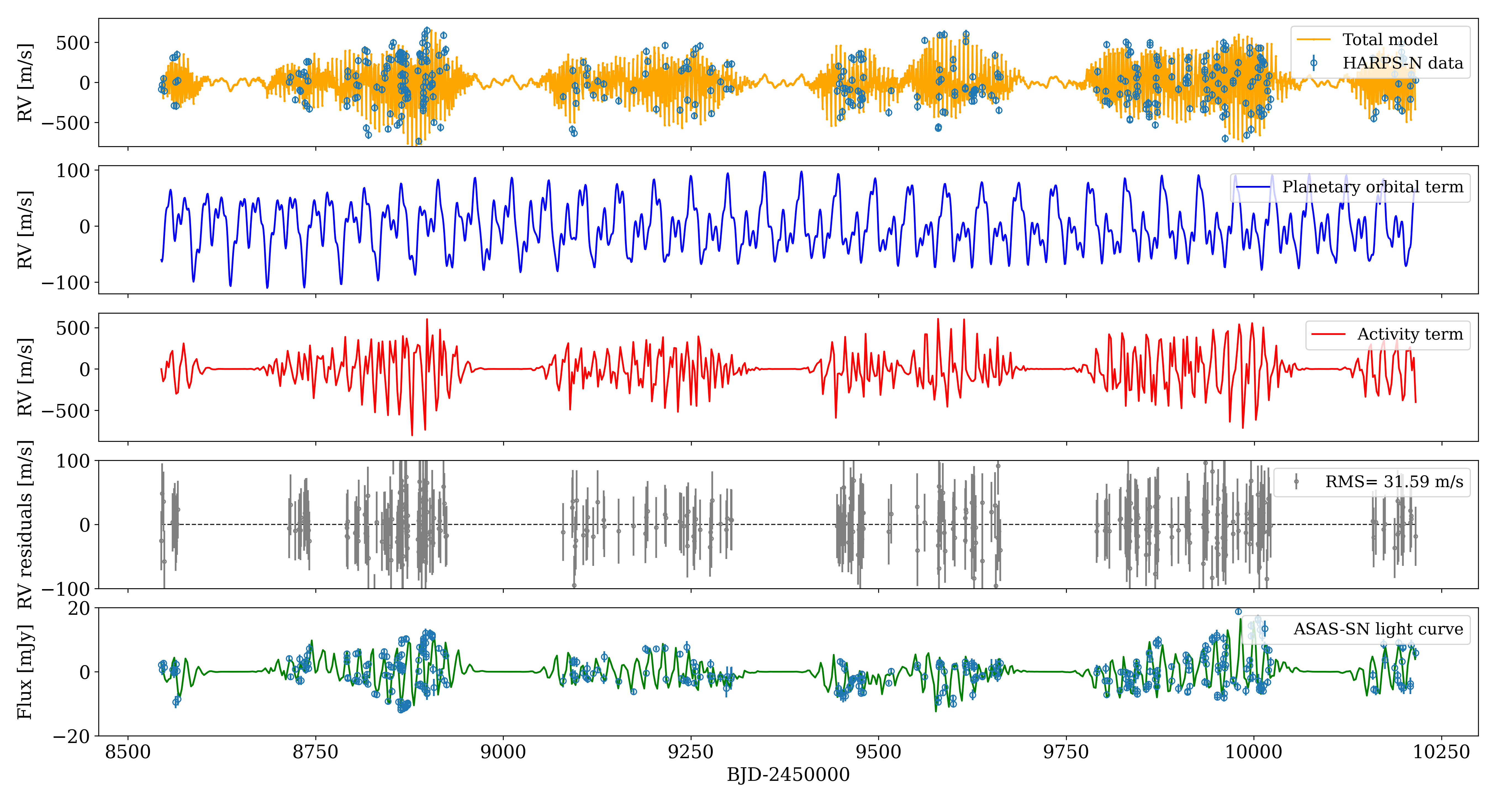}
    \caption{Results for the GP 2D regression using the ASAS-SN light curve as the activity diagnostic applied to \texttt{snfit} RVs. \textit{First panel.}  RV time series and total best-fit model planetary+activity signals (orange curve)  \textit{Second panel.} Best-fit model for the planetary orbital term only. \textit{Third panel.} Best-fit model for the GP QP activity term only \textit{Fourth panel.} RV residuals, after subtracting the total best-fit model from the \texttt{snfit} RV dataset. The error bars include a jitter term added in quadrature. \textit{Last panel.} Best-fit model for the ASAS-SN light curve.}
    \label{fig:resultssnfitasas}
\end{figure*}

The best-fit results for the planetary masses obtained for each test model and RV extraction method (given as the 15.87$^{\rm th}$, 50$^{\rm th}$, and 84.14$^{\rm th}$ percentiles of the posterior distributions) are summarised in Table \ref{tab:massresults} and Fig. \ref{fig:summarymasses}. The marginalised posteriors for the six individual models are shown in Fig. \ref{fig:masses1D}--\ref{fig:masses3DASASCRX}. In a few cases, we obtained measurements with a significance greater than 3$\sigma$ (0.135$^{\rm th}$--99.865$^{\rm th}$ percentile range) for planets b, c, and e, but there is no single test model which gives well constrained results for all the masses at the same time. The mass of V1298\,Tau\,b is found in the range $\sim$0.5--0.7 $M_{\rm Jup}$ when the activity is modelled with a GP 1D regression, both using the \texttt{serval} and \texttt{snfit} dataset, and with a GP 2D approach (activity diagnostic: H$\alpha$ index and ASAS-SN light curve), in this case only for the RVs extracted with \texttt{snfit}. The mass of V1298\,Tau\,c is significantly constrained around $\sim$0.1 $M_{\rm Jup}$ only for the RVs extracted with \texttt{serval}, and using a multi-dimensional GP approach (2D, with the ASAS-SN photometry; 3-dim, with the ASAS-SN photometry and CRX index). For V1298\,Tau\,e, a $\geq$3$\sigma$ significant mass is measured in three cases in the range 0.4--0.7 $M_{\rm Jup}$ using a multi-dimensional GP regression: 2D, with the H$\alpha$ index or ASAS-SN photometry (\texttt{snfit}), and 3-dim, with the ASAS-SN photometry and CRX index (\texttt{serval}). The GP 2D models with the H$\alpha$ index or ASAS-SN photometry as activity diagnostic, give similar results when applied to the \texttt{snfit} dataset, and are those providing >3$\sigma$ significant masses for b and e at the same time. 
Limited to the cases discussed above, we obtain a significant 3$\sigma$ eccentricity for planet b for the case with a GP 1D regression and \texttt{serval} RVs (e=0.16$\pm$0.05; this represents the value at the start epoch of the $N$-body dynamic integration).
We show in Fig. \ref{fig:resultssnfitasas} the best-fit solution corresponding to the \texttt{snfit} dataset and a 2D GP regression analysis using the ASAS-SN photometry (>3$\sigma$ significant masses for planets b and e), as an example of our results. The best-fit values for all the free-parameters of the model are summarised in Table \ref{tab:results2dasas}. In Fig. \ref{fig:singleplanets} we show the RV signals due to each individual planet. In Appendix \ref{sec:overfitting} we examine the possibility for overfitting for this specific model, by following a cross-validation (CV) test as described by \cite{Blunt2023AJ....166...62B}, that was applied to the data of V1298\,Tau analysed by \cite{Suarez2022NatAs...6..232S}. That analysis shows that the model is overfitting the data, given that its predictive capability decreases as the sampling rate
approaches $P_{\rm rot}$. In Appendix \ref{sec:stabilityanalysis}, we use the posteriors of the planetary parameters to explore the dynamical stability of the system over 5 Myr.

\subsubsection{Model comparison analysis using MGIC$_{\rm RV}$}
Recently, \cite{barragan26} introduced a new statistical criterion for comparing models based on multi-dimensional GPs. The method consists of calculating the MGIC$_{\rm RV}$ indicator (defined by Equation 17 in \citealt{barragan26}) for each model, and evaluating the difference $\Delta$(MGIC$_{\rm RV}$) between pairs of models to determine which one better represents the time variability of the RV component. \cite{barragan26} proposes a metric according to which models with $\Delta$(MGIC$_{\rm RV}$)$\lesssim$10 are considered statistically equivalent, whereas a larger difference indicates a preference for the model characterized by the lower MGIC$_{\rm RV}$ value, although this threshold should be treated as indicative. We derived MGIC$_{\rm RV}$ for each model and RV extraction method, by adopting the maximum likelihood parameters for the calculation. We report the $\Delta$(MGIC$_{\rm RV}$) values in Table \ref{tab:massresults}, evaluated using the 1D GP model as the reference model. According to this criterion, the 1D GP model appears clearly statistically inadequate to describe the RV variability, and the model using the BIS is strongly favoured over all others. This model does not yield a detection at the 3$\sigma$ significance level for any of the four planets. This implies that, assuming the \cite{barragan26} criterion is robust also for the case of V1298\,Tau, the detections obtained with other activity indicators are spurious. Furthermore, in the case of RVs extracted with \texttt{serval}, the CRX index and ASAS-SN photometry are activity indicators with comparable performance, while for RVs extracted with \texttt{snfit}, photometry and the H$\alpha$ index have equivalent relevance.

Since V1298\,Tau is one of the very first complex cases to which the MGIC$_{\rm RV}$-based criterion is applied, we consider it appropriate to leave the detailed exploration of the physical meaning of the model comparison results for a future work, specifically to understand why the BIS proves to be the most effective stellar activity diagnostic compared to the others we tested, and what would produce spurious $3\sigma$ detections in our modelling framework. As a final remark, the GP hyper-parameters $\theta$, $\tau$, and $w$, which are shared in the GP kernels used for both the RV and activity diagnostic datasets, are well constrained for each test model (Fig. \ref{fig:hyperpost}).  

\begin{table*}
    \centering
    \tiny
        \caption{Planetary mass measurements for the models tested in this work. }
    \begin{tabular}{lcccccc}
    \hline\hline
    \noalign{\smallskip}
     RV dataset; activity model & & \multicolumn{4}{c}{Mass $M_p^{(a)}$} & $\Delta$(MGIC$_{\rm RV}$)$^{(b)}$\\
     \hline
     \noalign{\smallskip}
      Planet & & b & c & d & e & \\
     \hline
      \noalign{\smallskip}
      \texttt{serval}; GP 1D & [$M_{\rm Jup}$] & 0.52$^{+0.12}_{-0.14}$ ($<0.87$)& 0.092$^{+0.030}_{-0.037}$ ($<0.18$)& 0.097$^{+0.042}_{-0.043}$ ($<0.21$)&  0.41$^{+0.17}_{-0.18}$ ($<0.88$) & 0\\
       \noalign{\smallskip}
       & [\mearth] & 164$^{+38}_{-45}$ & 29$^{+10}_{-12}$ & 31$^{+13}_{-14}$ &  130$^{+54}_{-59}$ \\
       \noalign{\smallskip}
        \texttt{serval}; GP 2D (BIS) & [$M_{\rm Jup}$] & 0.18$\pm$0.12 ($<0.51$) & 0.071$^{+0.036}_{-0.041}$ ($<0.18$) & 0.046$^{+0.038}_{-0.031}$ ($<0.17$) &  0.18$^{+0.11}_{-0.10}$ ($<0.50$) & -331.5 \\
       \noalign{\smallskip}
      & [\mearth] & 58$^{+40}_{-38}$ & 23$^{+11}_{-13}$ & 15$^{+12}_{-10}$ &  58$^{+35}_{-33}$ \\
          \noalign{\smallskip}
      \texttt{serval}; GP 2D (CRX index) & [$M_{\rm Jup}$] & 0.30$^{+0.14}_{-0.16}$ ($<0.36$) & 0.097$^{+0.039}_{-0.037}$ ($<0.21$) & 0.045$^{+0.037}_{-0.029}$ ($<0.16$) &  0.31$^{+0.12}_{-0.14}$ ($<0.64$) & -206.3 \\ 
     \noalign{\smallskip}
      & [\mearth] & 96$^{+45}_{-52}$ & 31$\pm$12 & 14$^{+12}_{-9}$ &  97$^{+37}_{-44}$ \\
      \noalign{\smallskip}
      \texttt{serval}; GP 2D (H$\alpha$ index) & [$M_{\rm Jup}$] & 0.34$^{+0.12}_{-0.15}$ ($<0.68$) & 0.077$\pm$0.036 ($<0.16$) & 0.048$^{+0.034}_{-0.029}$ ($<0.15$) &  0.25$\pm$0.11 ($<0.57$) & -160.4 \\ 
       \noalign{\smallskip}
       & [\mearth] & 107$^{+37}_{-47}$ & 24$\pm$11 & 15$^{+11}_{-9}$ &  79$^{+34}_{-36}$ \\
        \noalign{\smallskip}
        \texttt{serval}; GP 2D (photometry) & [$M_{\rm Jup}$] & 0.143$^{+0.137}_{-0.095}$ ($<0.53$) & 0.107$\pm$0.035 ($<0.22$) & 0.039$^{+0.038}_{-0.026}$ ($<0.16$) &  0.34$^{+0.15}_{-0.16}$ ($<0.89$) & -201.2 \\ 
       \noalign{\smallskip}
      & [\mearth] & 45$^{+44}_{-30}$ & 34$\pm$11 & 12$^{+12}_{-8}$ &  109$^{+49}_{-50}$ \\
        \noalign{\smallskip}
          \texttt{serval}; GP 3-dim (photom. $\&$ CRX index) & [$M_{\rm Jup}$] & 0.22$^{+0.17}_{-0.14}$ ($<0.60$) &  0.117$\pm$0.033 ($<0.22$) & 0.050$^{+0.050}_{-0.035}$ ($<0.17$) & 0.40$^{+0.14}_{-0.13}$ ($<0.78$) & -290.6 \\ 
       \noalign{\smallskip}
       & [\mearth] & 69$^{+53}_{-44}$ & 33$\pm$10 & 16$^{+16}_{-11}$ & 127$\pm$44  \\
       \noalign{\smallskip}
      \hline
      \noalign{\smallskip}
             \texttt{snfit}; GP 1D & [$M_{\rm Jup}$] & 0.65$^{+0.21}_{-0.18}$ ($<1.28$) & 0.11$\pm$0.07 ($<0.29$) & 0.10$^{+0.10}_{-0.06}$ ($<0.37$) &  0.58$^{+0.23}_{-0.24}$ ($<1.27$) & 0 \\
       \noalign{\smallskip}
      & [\mearth]  & 206$^{+66}_{-59}$ & 36$^{+23}_{-22}$ & 33$^{+31}_{-20}$ &  186$^{+73}_{-77}$ \mearth\\
\noalign{\smallskip}
\texttt{snfit}; GP 2D (BIS) & [$M_{\rm Jup}$] & 0.23$^{+0.16}_{-0.14}$ ($<0.68$) & 0.057$^{+0.051}_{-0.038}$ ($<0.22$) & 0.068$^{+0.067}_{-0.046}$ ($<0.28$) &  0.29$^{+0.17}_{-0.16}$ ($<0.78$) & -416.7 \\
     \noalign{\smallskip}
     & [\mearth] & 73$^{+49}_{-45}$ & 18$^{+16}_{-12}$ & 22$^{+21}_{-15}$ &  92$^{+53}_{-51}$ \\
\noalign{\smallskip}    
        \texttt{snfit}; GP 2D (H$\alpha$ index) & [$M_{\rm Jup}$] & 0.78$^{+0.19}_{-0.20}$ ($<1.50$)  & 0.079$^{+0.059}_{-0.052}$ ($<0.25$)  & 0.064$^{+0.18}_{-0.064}$ ($<0.24$)  &  0.66$^{+0.17}_{-0.18}$ ($<1.17$) & -119 \\
         \noalign{\smallskip}
         & [\mearth] & 249$^{+61}_{-63}$ &  25$^{+19}_{-17}$ & 20$^{+17}_{-12}$ & 211$^{+54}_{-57}$ \\
          \noalign{\smallskip}     
      \texttt{snfit}; GP 2D (photometry) & [$M_{\rm Jup}$] & 0.67$^{+0.16}_{-0.19}$ ($<1.15$)  & 0.18$^{+0.07}_{-0.08}$ ($<0.38$)  & 0.17$^{+0.07}_{-0.08}$ ($<0.37$)  &  0.71$^{+0.18}_{-0.22}$ ($<1.31$) & -126.4 \\
         \noalign{\smallskip}
       &  [\mearth] & 213$^{+50}_{-62}$ &  57$^{+21}_{-25}$ & 54$^{+22}_{-27}$ & 225$^{+57}_{-70}$ \\
          \noalign{\smallskip}
          \hline
    \end{tabular}
    \tablefoot{
    \tablefoottext{a}{The values are given as the 15.87$^{\rm th}$, 50$^{\rm th}$, and 84.14$^{\rm th}$ percentiles of the posterior distributions, and are reported both in units of Jupiter and Earth masses. The upper limits corresponding to the 99.865$^{\rm th}$ percentile (3$\sigma$) are given in parenthesis for masses expressed in Jupiter mass units, and used in Fig. \ref{fig:summarymasses}. The mass of the host star was fixed to 1.170 \msun~ \citep{Suarez2022NatAs...6..232S}.}
    \tablefoottext{b}{$\Delta$MGIC$_{\rm RV}$ is calculated with reference to the GP 1D model, i.e. $\Delta$(MGIC$_{\rm RV}$)=MGIC$_{\rm RV,\,N-dim}$-MGIC$_{\rm RV,\,1D}$, with $N$=2,3).}
    }
    \label{tab:massresults}
\end{table*}

\subsection{injection and recovery simulations assuming planets' masses as measured by L26} \label{sec:injret}

\begin{figure*}
    \centering
    \includegraphics[width=0.9\textwidth]{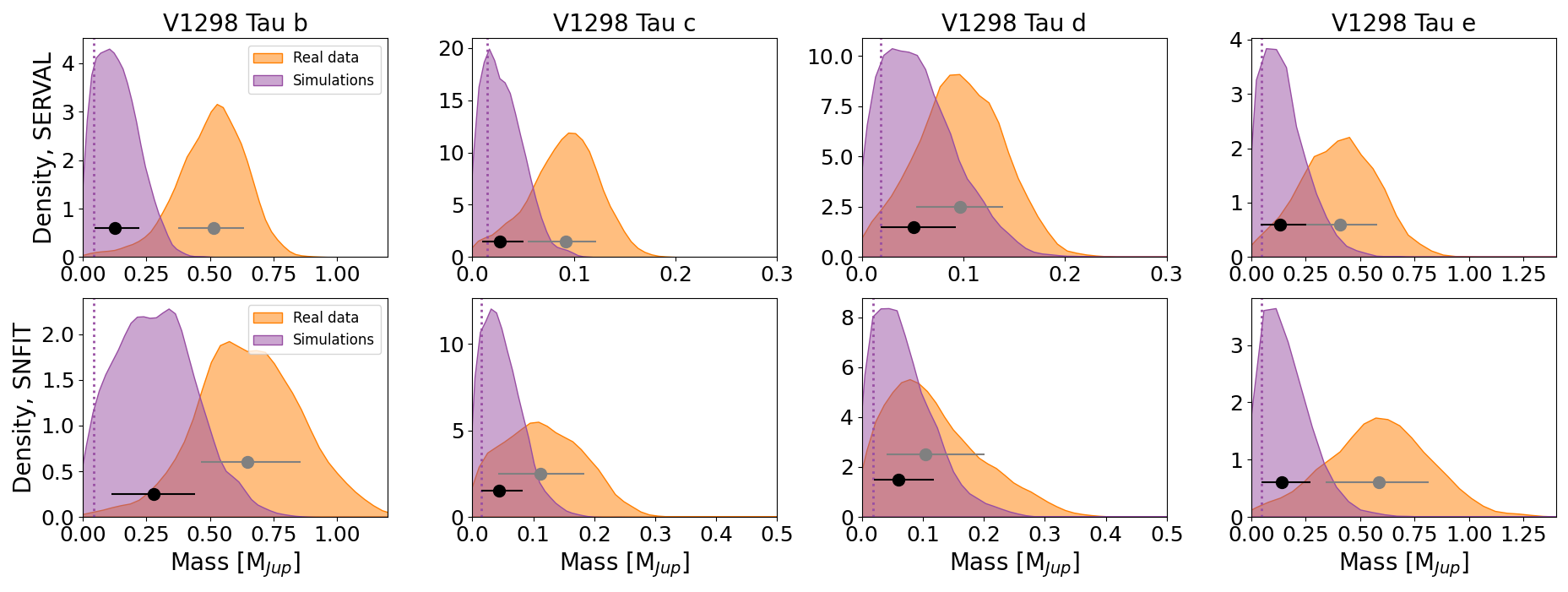}
    \caption{Comparison between the posterior distributions for the masses of the four confirmed transiting planets in the V1298\,Tau, for real (violet) and simulated (orange) data. \textit{First row.} Results for \texttt{serval} RVs. \textit{Second row.} Results for \texttt{snfit} RVs. These results refer to the case of a 1D GP QP regression. The median and the percentile range 15.87$^{\rm th}$--84.14$^{\rm th}$ (1$\sigma$) are indicated as a black and grey symbols, for the simulated and real cases, respectively. Dotted vertical lines identify the injected planetary masses, based on the results of L26. }
    \label{fig:masses1Dsimulation}
\end{figure*}

\begin{figure*}
    \centering
    \includegraphics[width=0.9\textwidth]{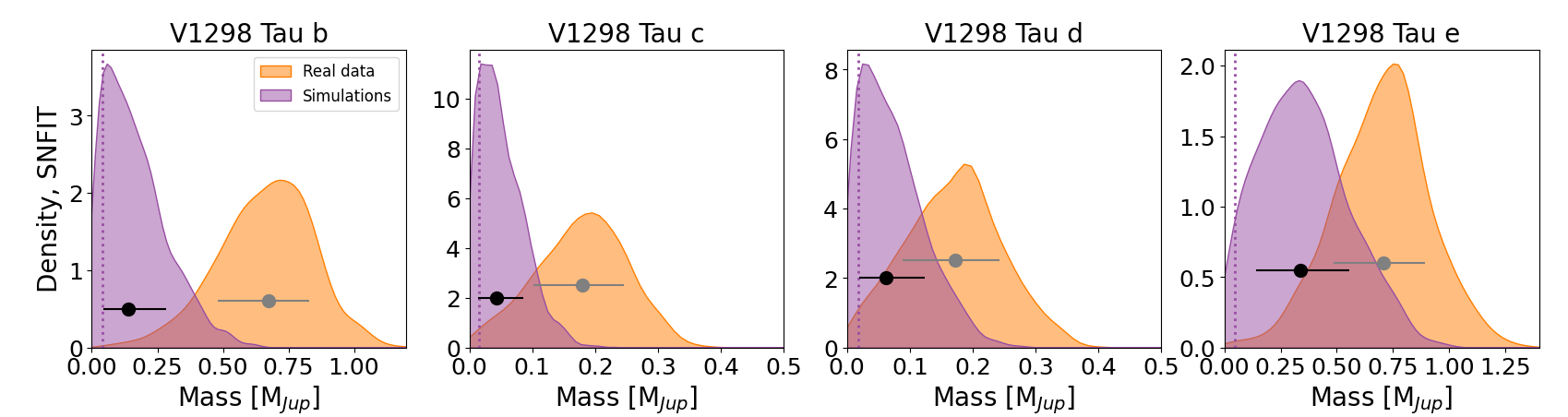}
    \caption{Same as in Fig. \ref{fig:masses1Dsimulation}, but for a GP 2D regression using the ASAS-SN light curve as activity diagnostic, applied to the RVs extracted with \texttt{snfit}.}
    \label{fig:masses2Dsimulation}
\end{figure*}

Through a TTV analysis, L26 measured masses of the four planets that are in the super-Earth/sub-Neptune range (Table \ref{Table:Summary_planets}). Part of the results of our study suggest that some of the planets could be more massive, therefore showing possible inconsistency between dynamical mass measurements obtained with two different techniques. To test our analysis framework against biased detections, we used injection and recovery simulations adopting the masses found by L26 as the real planetary masses. The idea is that by simulating RV Doppler variations due to the low-mass planets identified by L26, if we get mass posteriors compatible with to those obtained from real data, which are indicative of more massive planets, that would suggest that the massive planets identified by our analysis framework are likely an artifact. Going into the details of the simulations, for our tests we considered the GP 1D model, analysing the RVs extracted with \texttt{serval} and \texttt{snfit}, and the GP 2D model with the ASAS-SN photometry used as activity tracer, in that case considering only the RVs calculated with \texttt{snfit} because of the >3$\sigma$ significant masses measured for planets b and e. The mock RV datasets are built by adopting the best-fit correlated activity signals found by modelling the real data in each case, to which we added the total Doppler signal due to the four planets from L26 simulated with \texttt{trades}. The best-fit parameters corresponding to these activity signals are in agreement with those obtained after fitting the real data with a model that includes only the GP kernel and no planets. This result shows that the RV variability modelled by a QP kernel can be assumed as a realistic representation of the stellar activity signal, and used as a template onto which to inject planetary signals. Finally, Gaussian noise was randomly added to each data point using an half-Gaussian distribution with zero mean and width equal to the sum in quadrature of $\sigma_{\rm RV}(t)$ and $\sigma_{\rm RV,\,jitt.}$. The values of the uncorrelated jitter that we determined in each case are (\textit{i}) $\sigma_{\rm RV,\,jitt.}$=17.4 and 33.4 \ms\, for the GP 1D model and \texttt{serval} and \texttt{snfit} RVs, respectively, and (\textit{ii}) $\sigma_{\rm RV,\,jitt.}$=43.7 \ms\, for the GP 2D model with \texttt{snfit} RVs. For the GP 2D case, for the photometric dataset we used the same ASAS-SN time series that was modelled with the real RV dataset. 
To fit the simulated RV time series, we adopted the same assumptions and modelling framework as described in Sect. \ref{sec:modellingframework}. Fig. \ref{fig:masses1Dsimulation} and \ref{fig:masses2Dsimulation} show the posterior distributions for the planets' masses obtained for the mock datasets, compared to the posteriors corresponding to the original RV datasets (also shown in Fig. \ref{fig:masses1D} and \ref{fig:masses2DASAS}). For the case with a GP 1D, none of the posterior distributions for the mock RV dataset correspond to a significant estimate (3$\sigma$) of the planetary mass, but only upper limits can be determined for all the planets. That is consistent with the low masses of the simulated planets. To assess whether the posterior distributions for real and simulated data, and for each planet, differ from one another, we calculated the ratio

\begin{equation}
  g= \frac{|\tilde{\theta}_{\rm real} - \tilde{\theta}_{\rm simulation}|}{\sqrt{\sigma_{\rm real}^2 + \sigma_{\rm simulation}^2}}
\end{equation}

where the parameters are defined as follows:
\begin{itemize}
    \item $\tilde{\theta}_{\rm real}, \tilde{\theta}_{\rm simulated}$: median values of the marginal posterior distributions.
    \item $\sigma_{\rm real}, \sigma_{\rm simulated}$: $1\sigma$ uncertainties corresponding to one of the asymmetric error bars, i.e. that on the side of a median facing the opposing median value. Specifically, since in our case it is always $\tilde{\theta}_{\rm simulated} < \tilde{\theta}_{\rm real}$, then $\sigma_{\rm simulated}$ represents the upper error of $\tilde{\theta}_{\rm simulated}$, and $\sigma_{\rm real}$ is the lower error of  $\tilde{\theta}_{\rm real}$.
\end{itemize}
The values of the ratio $g$ are summarised in Table \ref{tab:injretrievalstatistics}. A value of $g$ between 2$\sigma$ and 3$\sigma$ indicates a moderate disagreement between two distributions. 
We can conclude that the posteriors for the real data moderately differ from those of the mock datasets for planet b, in two cases: GP 1D (\texttt{serval}) and GP 2D (\texttt{snfit}). This result suggests that, at least limited to these cases, the mass of planet b measured from real data may be a significant fraction of Jupiter’s mass. For planet e, the simulations do not provide enough statistical support that it has a significant fraction of the Jupiter mass, and for planet c they do not provide evidence for $m_{\rm c}\sim$10--20\% $M_{\rm Jup}$. 

\begin{table}[]
    \centering
     \caption{Statistical tension calculated from  Eq. (3) for the planetary mass posterior distributions obtained from injection-and-recovery simulations. }
    \begin{tabular}{lcccc}
    \hline\hline
        Dataset and model & \multicolumn{4}{c}{Planet} \\
    \hline 
    & b & c & d & e \\
    \hline
        \texttt{serval}; GP 1D & 2.29 & 1.46 & 0.76 & 1.26 \\ 
        \texttt{snfit}; GP 1D & 1.51 & 0.87 & 0.52 & 1.61 \\ 
        \texttt{snfit}; GP 2D with photometry & 2.22 & 1.53 & 1.05 & 1.19 \\ 
        \hline
    \end{tabular}
    \label{tab:injretrievalstatistics}
\end{table}

\section{Conclusions} \label{sec:conclusions}

In this work, we presented a new determination of the dynamical mass of the four confirmed transiting planets in the infant V1298\,Tau system. Our analysis is based on nearly 400 spectroscopic observations with HARPS-N covering more than 4.5 years, which represents the largest and most homogeneous dataset available for this target. To account for the very high levels of stellar activity and planet-planet gravitational interactions, we adopted a modelling approach by coupling $N$-body integration with multi-dimensional GP regression, testing two different RV extraction methods and several activity diagnostic.

The main findings and caveats of our study can be summarised as follows:
\begin{itemize}
    \item \textit{Sensitivity to modelling choices.} The resulting planetary masses depend on the RV extraction pipeline (\texttt{serval} vs. \texttt{snfit}) and the selected activity diagnostics. This is certainly no surprise for a very complex target such as V1298\,Tau, but these results represent a useful starting point for more complex analyses.
    \item \textit{Planet b.} In some test cases we measured a $\geq$3$\sigma$ significant mass for planet b in the range $m_{\rm b}\sim$0.5-0.7 $M_{\rm Jup}$. Injection-and-recovery simulations applied to specific models indicate that the constraints for $m_{\rm b}$ are moderately significant. This finding appears to warrant attention, and it highlights a tension with the sub-Neptune/super-Earth regime suggested by recent TTV-only analyses. This discrepancy would show the potential limitations when only one of the two techniques is applied to this challenging system. Our study strengthen the awareness that the analysis of the RVs is drastically hampered by stellar activity for a target such as V1298\,Tau. For the case of TTVs, the identification of unique solutions can be affected by poor sampling, that prevent the identification of short-timescale variations, which is a necessary (although not sufficient) condition to avoid degeneracies \citep[i.e. solutions that differ in terms of planetary parameters such as masses, eccentricities, and orbital inclinations; e.g.][]{Lammers_2026}.   
    \item \textit{Planets c, d, and e.} Concerning V1298\,Tau c, we could place mass upper limits in most of the test models, and in two cases (using \texttt{serval} RVs) we measured a mass of $\sim$30 \mearth\, with $\sim$3$\sigma$ significance. For V1298\,Tau d, we could only determine mass upper limits in all tested models. For planet e, the 3$\sigma$ significant solutions ($m_{\rm e}\sim$0.4-0.7 $M_{\rm Jup}$), indicating a mass larger than that of a sub-Neptune, could be overestimated, as suggested by injection-and-recovery simulations.  
    \item \textit{Benchmarking the $\text{MGIC}_{\rm RV}$ model comparison criterion.} Applying the $\text{MGIC}_{\rm RV}$ metric \citep{barragan26} to the complex V1298\,Tau dataset shows that 1D GP models fail to describe the RV variability, whereas multi-dimensional GP models including BIS are strongly favoured, though yielding no $3\sigma$ planetary detections. According to $\text{MGIC}_{\rm RV}$ metric, 3$\sigma$ significant solutions found using other indicators (CRX, ASAS, $\text{H}_\alpha$, depending on the RV extraction pipeline) could be spurious. As V1298\,Tau represents one of the very first applications of this criterion, our work serves as a reference case study for future efforts evaluating the robustness of $\text{MGIC}_{\rm RV}$ in multi-planet infant systems with RV time series heavily dominated by stellar activity variability, and investigating the physical drivers of indicator effectiveness. In this paper, we refrain from drawing definitive conclusions based solely on $\text{MGIC}_{\rm RV}$, as evaluating whether this metric yields realistic estimates, or is subject to potential biases in extreme activity regimes, requires further systematic testing on larger samples. 
\end{itemize}

In conclusion, our mass measurements based on RVs are not precise enough to draw firm conclusion on the nature of the planets, but they do not leave out the possibility that some of these infant planets may be giant-like rather than low-density progenitors of sub-Neptunes, also taking the mass upper limits into account. If all the planets fall into the sub-Neptune/super-Earth category, as found by the analysis of TTVs alone (L26) or atmospheric transmission spectroscopy \citep{barat_2024A&A...692A.198B,Barat2024NatAs.tmp...86B}, not unexpectedly our dataset would not allow us to measure their masses, but just to determine upper limits, as we tested through injection and recovery simulations. 
A clearer picture will require the combination of RVs and photometry in a comprehensive photo-dynamical model, incorporating a larger sample of transits and exploiting the results of this work. As a final remark, it is important to consider that the architecture of the V1298\,Tau system may be even more complex than currently assumed. The possible presence of additional, non-transiting or unconfirmed companions could introduce further gravitational perturbations, affecting both TTV and RV signals. 
There is currently no study comparing masses determined via RV and TTV for planets in a multi-planetary system as young as V1298\,Tau, which is still dynamically ``hot''. For this reason, the results presented in this paper, obtained from the RV-only analysis, appear significant when compared with those obtained using other techniques, allowing for an investigation to identify the cause of biases. The conclusions that can be drawn may be of assistance in the study of infant multi-planetary systems close to the mean motion resonance, such as V1298\,Tau. 

\begin{acknowledgements}
We thank the anonymous referee for their useful comments, and O. Barrag\'an for useful discussions about data modelling and application of the model comparison analysis based on $\text{MGIC}_{\rm RV}$. Our work is based on observations made with the Italian Telescopio Nazionale \textit{Galileo} (TNG) operated by the Fundaci\'on Galileo Galilei (FGG) of the Istituto Nazionale di Astrofisica (INAF) at the Observatorio del Roque de los Muchachos (La Palma, Canary Islands, Spain). We acknowledge the Italian center for Astronomical Archives (IA2, \url{https://www.ia2.inaf.it}), part of the Italian National Institute for Astrophysics (INAF), for providing technical assistance, services and supporting activities of the GAPS collaboration. LBo acknowledges the funding support from Italian Space Agency (ASI) regulated by ``Accordo ASI-INAF n. 2013-016-R.0 del 9 luglio 2013 e integrazione del 9 luglio 2015 CHEOPS Fasi A/B/C''.
ASM acknowledges financial support from the Spanish Ministry of Science and Innovation (MICINN) projects PID2020-117493GB-I00, and 2024 Ramón y Cajal program MICINN RYC2024-050707-I, and from the Government of the Canary Islands project ProID2020010129. SD gratefully acknowledges support from the ``Programma di Ricerca Fondamentale INAF 2023'' of the National Institute of Astrophysics (Large Grant 2023 NextSTEPS)
\end{acknowledgements}

%
   \bibliographystyle{aa} 
   \bibliography{V1298tau} 
%
\begin{appendix}

\section{Analysis of the GP predictive performance through a cross validation test} \label{sec:overfitting}
\cite{Blunt2023AJ....166...62B} highlighted the risks due to data overfitting in the case of GP-based complex models applied to V1298\,Tau RVs by questioning the reliability of the masses estimated by \cite{Suarez2022NatAs...6..232S}. They used a definition for overfitting as the inability of a model to predict data at new -or out-of-sample- epochs, and assessed it through an implementation of a CV test. In the case of a GP regression, if the adopted GP kernel primarily models stochastic noise instead of catching the astrophysical correlated signals due to stellar activity, its predictive efficiency at unseen epochs gets worse. To investigate the possibility for overfitting in our analysis, we performed a CV test following the same approach as \cite{Blunt2023AJ....166...62B}. We considered the case where \texttt{snfit} RVs are modelled through a GP 2D regression using the ASAS-SN light curve as activity diagnostic, because this is one of the cases for which we measured significant masses of planets b and e, and it looks as an appropriate model for testing the robustness of our framework. For this model we obtained a best-fit uncorrelated jitter $\sigma_{\rm RV,\,jitt}$ of 43$\pm$4 \ms ($\sim$2.5 times the median internal error $\sigma_{\rm RV}$), and that the RV residuals show an RMS of 32 \ms.

\begin{figure*}
    \centering
    \includegraphics[width=0.9\textwidth]{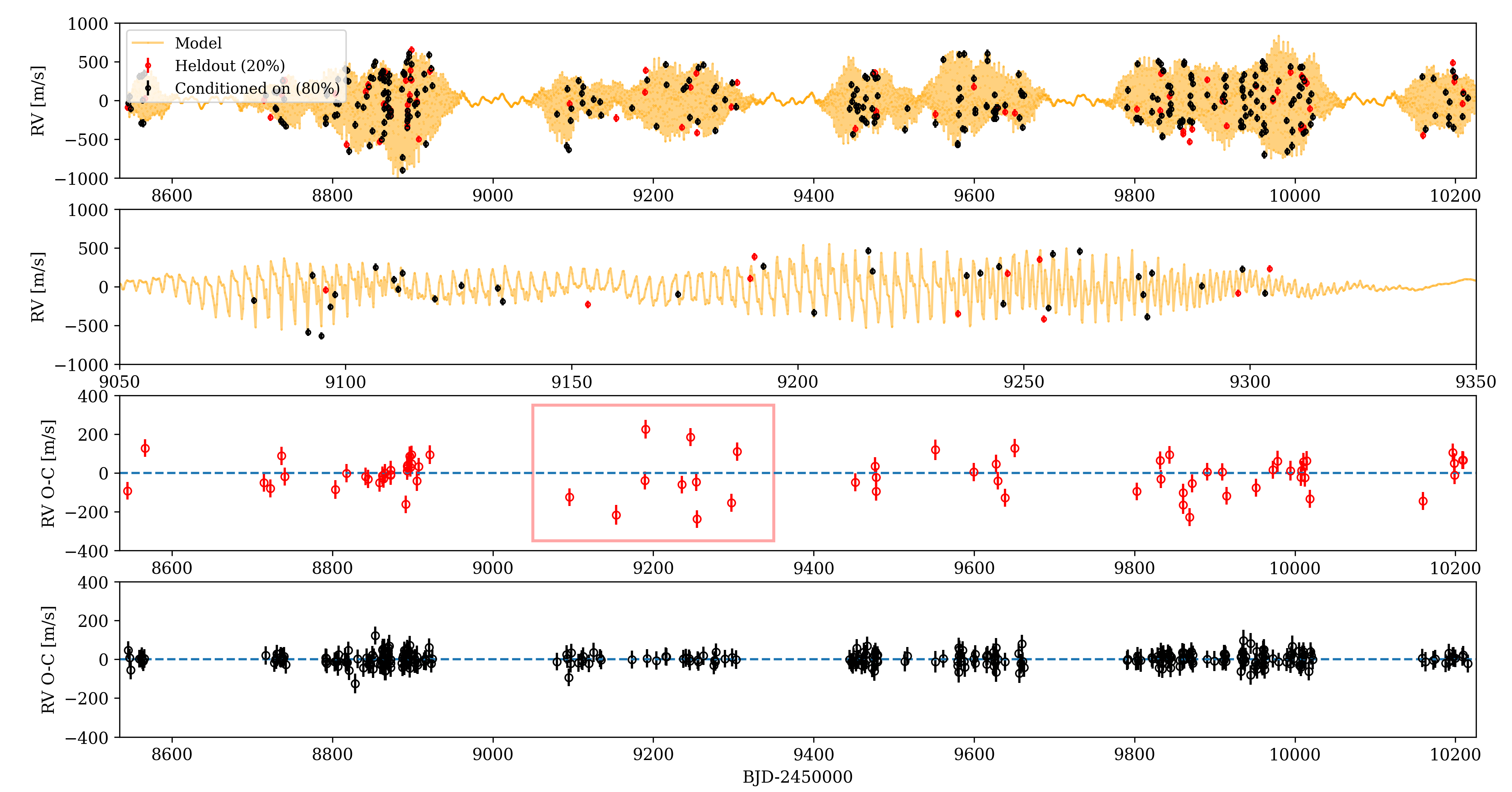}
    \caption{Example of one realisation of the CV test. The first panel shows the RV dataset, with the held-out and conditioned on data shown as red and black dots, respectively, and the best-fit model, trained on the conditioned on data, represented by an orange line. The second panel shows a section of the first panel for a better readability. The third and fourth panels show the residuals O-C of the two datasets. The red rectangle in the third plot identifies the time interval shown in the second panel. }
    \label{fig:CVexample}
\end{figure*}

\begin{figure}
    \centering
    \includegraphics[width=0.5\textwidth]{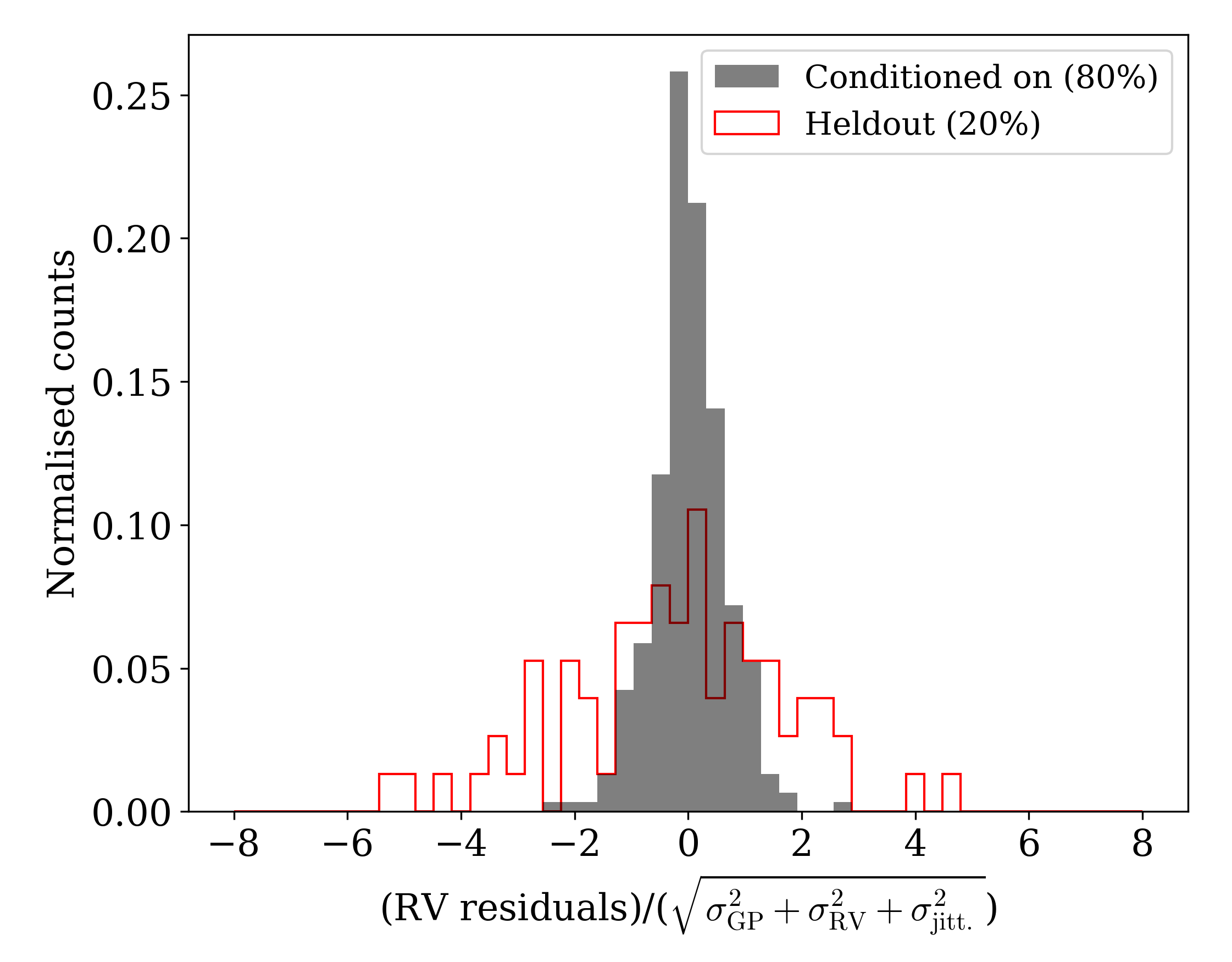}
    \caption{Distributions of the RV residuals O-C shown in the last two panels of Fig. \ref{fig:CVexample}, for the held-out and conditioned on subsamples. The O-C data are normalised to the total error budget.}
    \label{fig:CVhistoexample}
\end{figure}

The CV test requires that a conditioned-on (or training), and a held-out (or validation) RV subsamples be randomly extracted from the original time series. In our case, they represent the 80$\%$ and 20$\%$ of the entire dataset, respectively. First, the conditioned-on sample is used to predict the total model (planet-induced variability plus the GP QP stellar activity signal) at the epochs of the held-out sample, which are unseen by the trained model. Then, following \cite{Blunt2023AJ....166...62B}, the RV model residual O-C time series (O: observed data; C: values computed from the model) of the training and held-out samples, normalised to the total error budget $\sqrt{\sigma_{\rm RV}^{2}(t)+\sigma_{\rm GP}^{2}(t)+\sigma_{\rm jitt.}^{2}(t)}$, are compared to see if their distributions differ significantly: a broader and more uniform distribution for the held-out sample would be considered as evidence for model overfitting. An example of our CV test realisation is illustrated in Fig. \ref{fig:CVexample} and \ref{fig:CVhistoexample}, where we show the RV residuals for the training and held-out samples, and the corresponding distributions, respectively. The distribution of the held-out sample is broader than that of the conditioned-on data, and this could be interpreted as an affect of the GP overfitting. A sub-sample of the held-out residuals which show the largest scatter (highlighted by a red rectangle in the third panel of Fig. \ref{fig:CVexample}), and populate the tails of the histogram of Fig. \ref{fig:CVhistoexample}, correspond to data that fall within a time interval with a sparse sampling (second panel of Fig. \ref{fig:CVexample}). In this time range, the predictive power of the GP is limited by larger temporal gaps between training observations compared to intervals with denser data sampling.
We suspect that the larger scatter exhibited by the held-out residuals primarily reflects the loss of predictive coherence of the QP kernel over temporal gaps exceeding the stellar rotation period, rather than an incorrect parameterization of the GP model itself.

\begin{figure}
    \centering
    \includegraphics[width=0.5\textwidth]{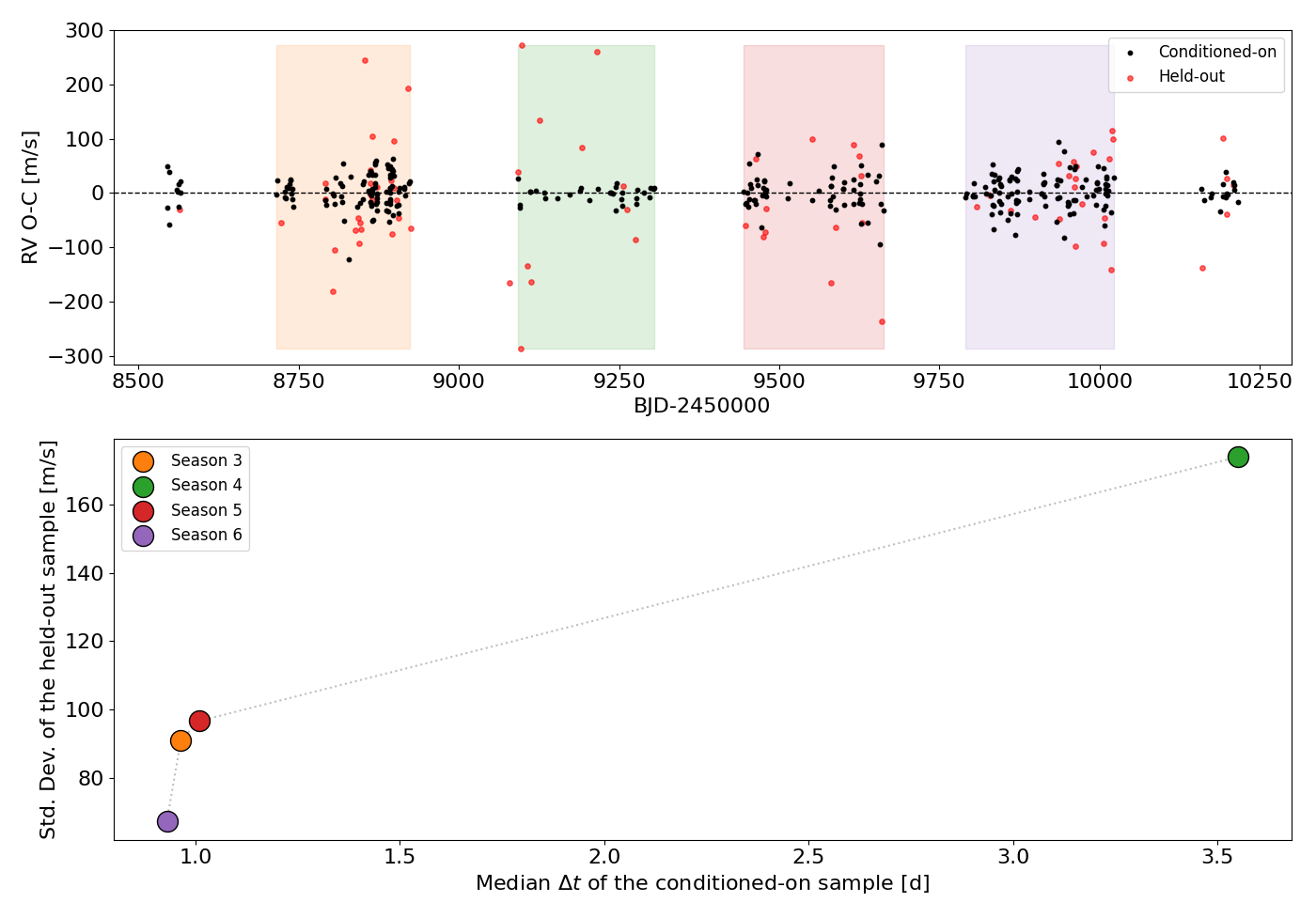}
    \caption{One example of the performed CV tests ($N$=200 random realisations. \textit{First panel.} Conditioned-on and held-out subsamples. The samples are analysed on a seasonal basis following the method described in Appendix \ref{sec:overfitting}. \textit{Second panel.} Standard deviation of the held-out subsamples in each observing season, as a function of the median temporal separation ($\Delta t$) between two consecutive observations in the conditioned-on dataset. Only seasons that contain at least 8 held-out data points have been considered in the statistical analysis. A colour code is used to indicate each data point and its corresponding seasonal subsample in the upper panel.}
    \label{fig:CVsamplingexample}
\end{figure}

\begin{figure}
    \centering
    \includegraphics[width=0.5\textwidth]{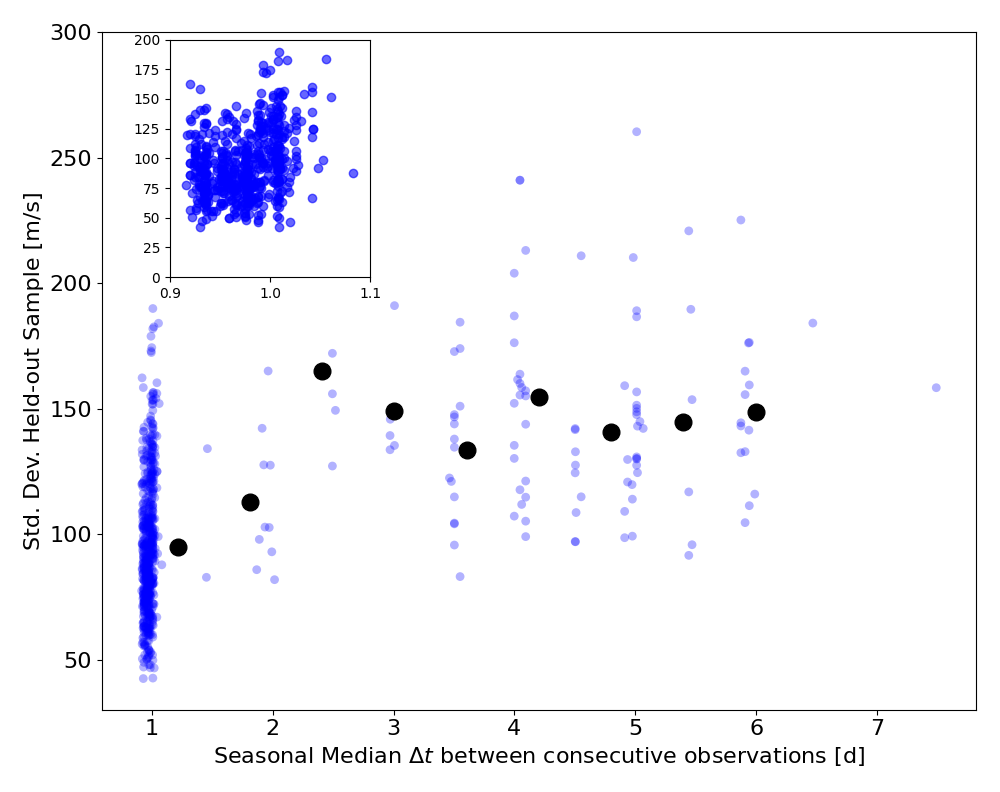}
    \caption{Results for the $N$=200 random realisations of the CV tests. The standard deviation of the held-out subsamples in each observing season (blue dots) is shown as a function of the median temporal separation ($\Delta t$) between two consecutive observations in the conditioned-on dataset. Black dots represent the mean values in bins of $\Delta t$. Only seasons that contain at least 8 held-out data points have been considered in the statistical analysis.}
    \label{fig:CVsamplingcumulative}
\end{figure}

To evaluate that, we performed a series of $N$=200 randomized CV tests. The data were analysed within time bins defined by observing seasons. To ensure statistical reliability, each bin was required to contain at least 8 held-out points. Within each bin, we characterised the sampling density by the median temporal separation ($\Delta t$) between two observations in the conditioned-on dataset, and investigated how the standard deviation of the residuals O-C in the held-out dataset changes as a function of ($\Delta t$). We show in Fig. \ref{fig:CVsamplingexample} an example of random realisation of the CV test. The plot shows that a higher $\Delta t$ corresponds to an increased scatter of the held-out residuals, and an explanation of this behaviour is that the GP model's ability to accurately predict values at unseen epochs degrades. 
Most of CV realisations in our sample show a similar correlation. To examine the statistical behaviour, we used the whole sample of $N$ realisations to produce a similar plot, shown in Fig. \ref{fig:CVsamplingcumulative}. In the high-cadence regime ($\Delta t \lesssim 1$ day), the GP effectively ``learns'' the underlying stellar signal, achieving a low residual scatter that represents the intrinsic noise floor of the data. The sharp increase at larger $\Delta t$ marks the threshold ($\Delta t\sim$3 days, corresponding to the stellar rotation period) where the model transitions from a physically representative state to an overfitted one. 
Fig. \ref{fig:CVsamplingcumulative} provides a quantitative measure of the model's overfitting, characterising how the predictive reliability at unobserved timestamps is affected by moving towards a sampling cadence longer than $P_{\rm rot}$. As the seasonal median $\Delta t$ increases, the GP loses the necessary temporal constraints to anchor the QP kernel. Consequently, while the model may achieve an excellent fit on the conditioned-on epochs, it fails to generalise to the held-out epochs.

\section{System dynamical stability} \label{sec:stabilityanalysis}
\begin{figure}[h]
    \centering
    \includegraphics[width=0.45\textwidth]{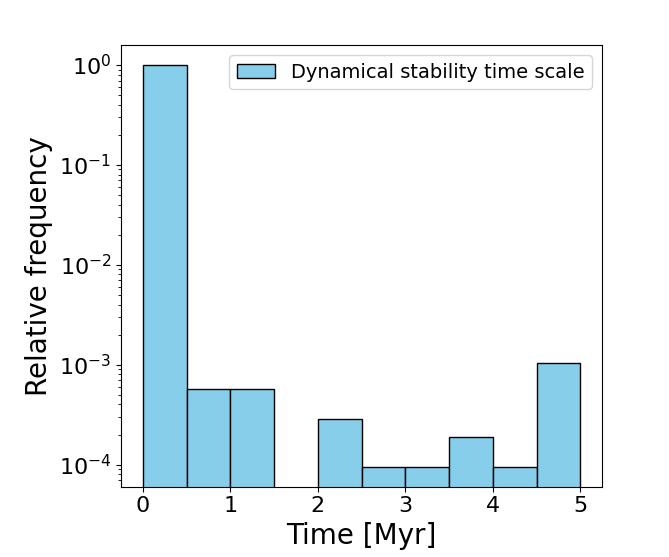}\\
    \includegraphics[width=0.45\textwidth]{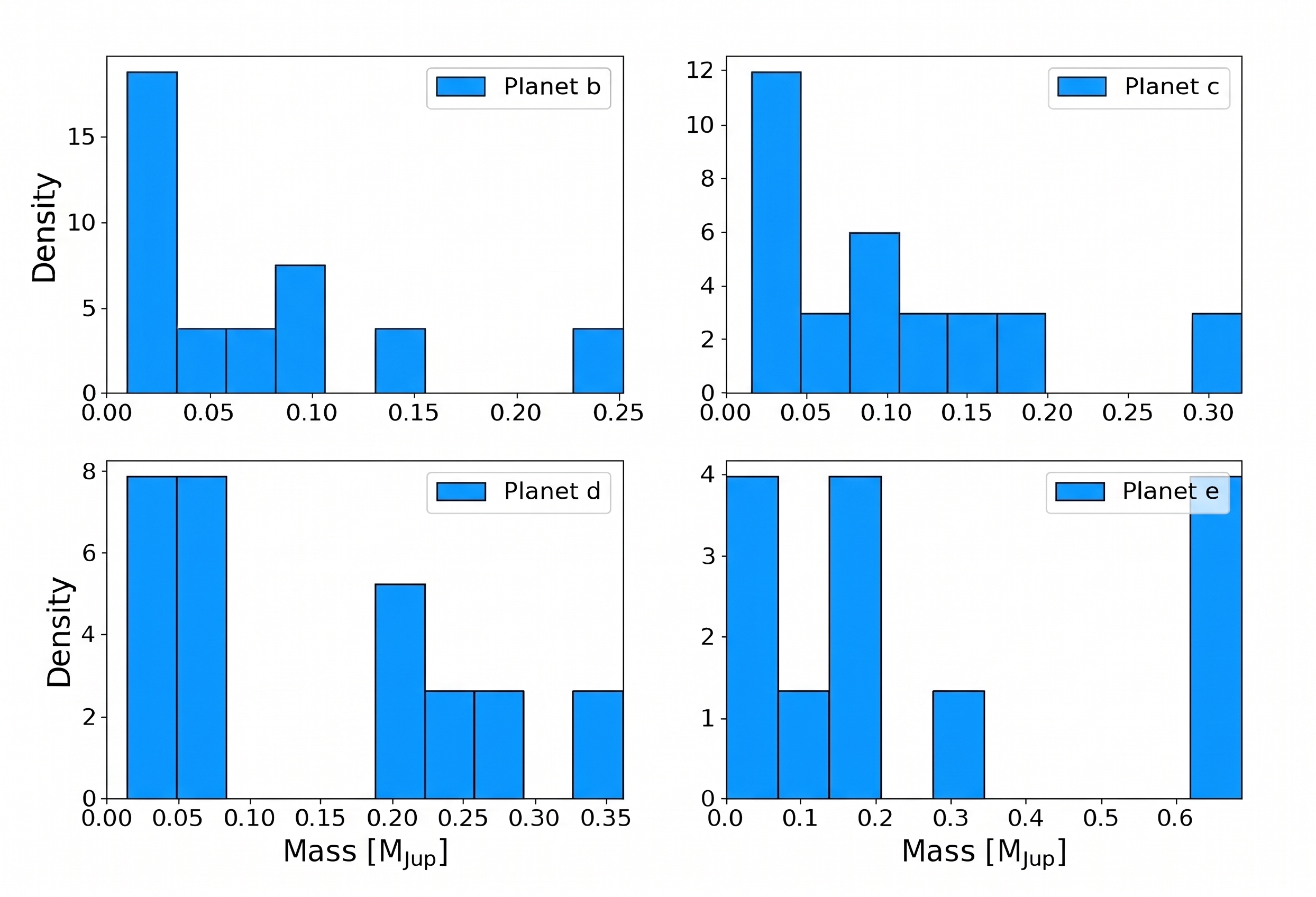}\\
    \caption{Results of the system dynamical stability analysis for the model GP 2D \texttt{snfit} with ASAS-SN photometry): distribution of the stability timescales (upper panel), and planet masses (lower plot) for the dynamically stable systems.}
    \label{fig:dynamicalanalysis}
\end{figure}

We explored the dynamical stability of the system by giving the orbital parameters and the relative masses of each MC posterior sample as input to the $N$-body integrator, letting the system evolve over a time span of 5 Myr. Even for this exercise we selected the case where \texttt{snfit} RVs are modelled through a GP 2D regression using the ASAS-SN light curve as activity diagnostic, and used the open source $N$-body code \texttt{rebound} \citep{rebound} because it provides the value of the chaos indicator MEGNO (Mean Exponential Growth of Nearby Orbits) at each step of the orbital integration (we verified that \texttt{rebound} gives results consistent with \texttt{trades}). The simulations were integrated using WHFast, a symplectic Wisdom-Holman integrator \citep{reboundwhfast,wh}, fixing a time step equal to $\sim$0.4 d (5\% of the orbital period of the innermost planet V1298\,Tau\,c). Variational equations were used to calculate trajectories of nearby orbits \citep{reboundvar}. 
We evaluated the orbital parameters at time steps of 500 years, and checked the stability against close encounters of consecutive pairs of planets (i.e. their distance must be less than the Hill radius), and verifying that the eccentricity of the orbit of each planet is less than 0.5. 
The results, summarised in Fig. \ref{fig:dynamicalanalysis}, show that only 0.1$\%$ of the posterior configurations are stable up to 5 Myr, with almost the total configurations surviving for less than 0.5 Myr.

\section{Additional tables and plots}

\begin{figure}[h]
    \centering
    \includegraphics[width=0.45\textwidth]{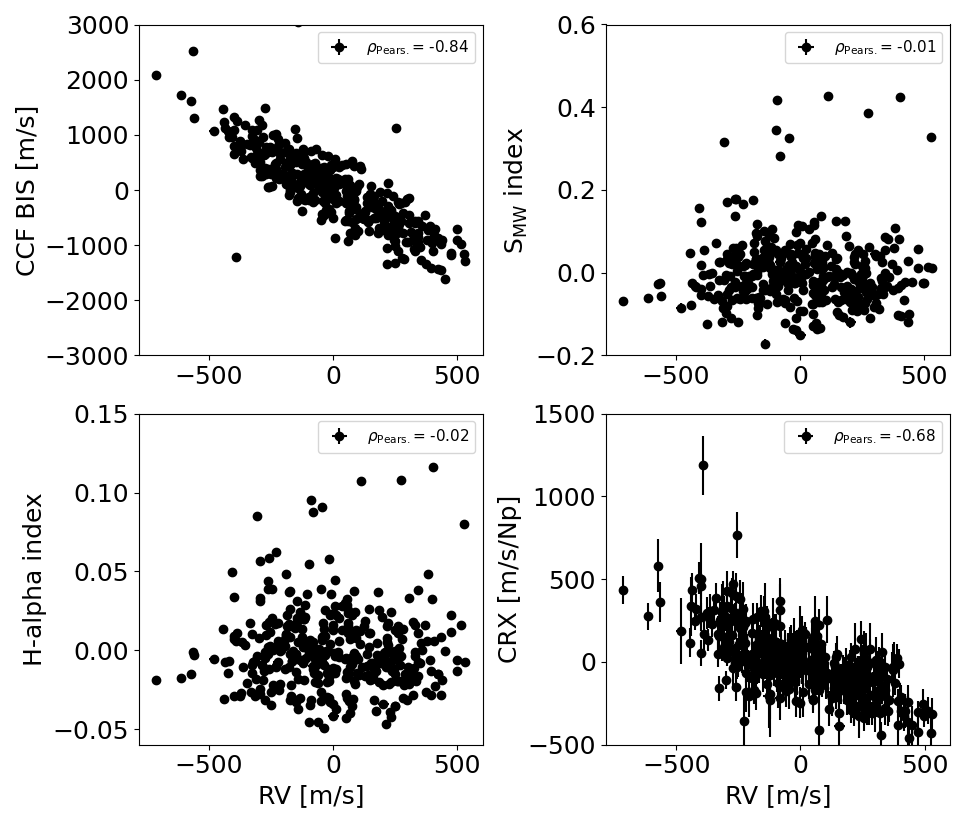}
    \caption{Correlation plots between \texttt{serval} RVs and the activity indicators BIS, $S_{\rm MW}$ index, H$\alpha$ index, and CRX. The Pearson's correlation coefficients $\rho_{\rm Pears.}$ are indicated within each panel. }
    \label{fig:rv_vs_act_indicators}
\end{figure}

\begin{figure}[h]
    \centering
    \includegraphics[width=0.435\textwidth]{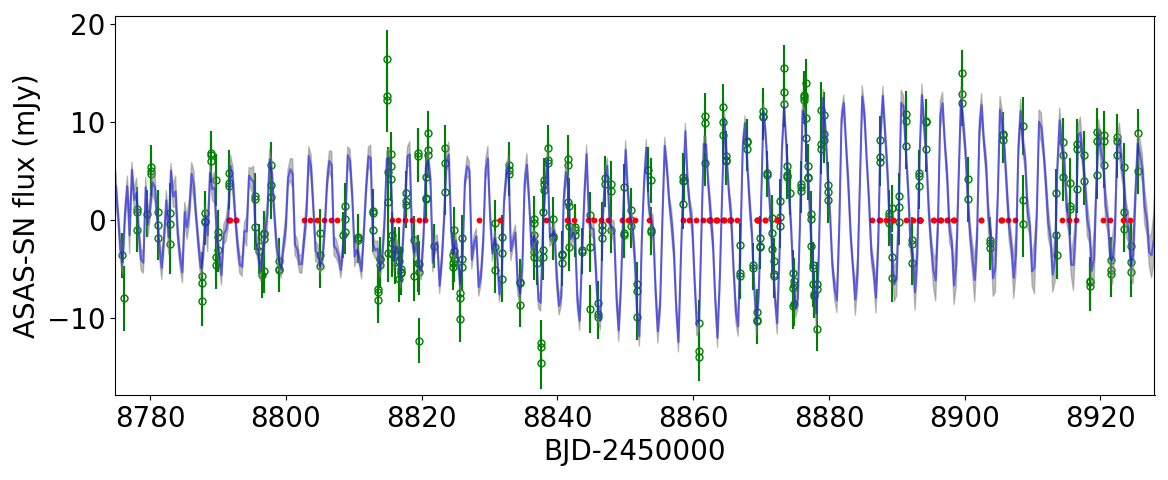}\\
    \includegraphics[width=0.45\textwidth]{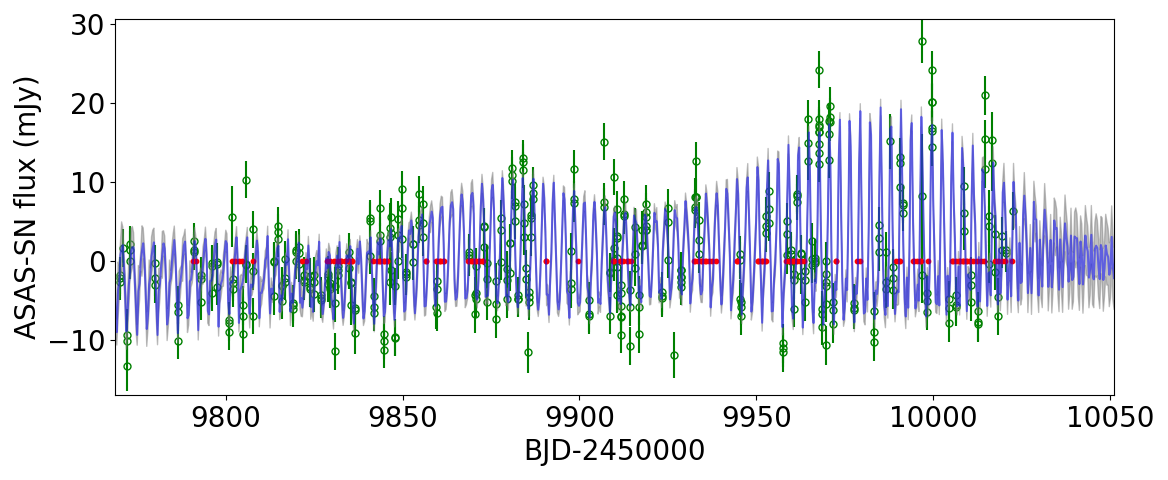}
    \caption{ASAS-SN \textit{g}' band light curve of V1298\,Tau (green dots; the mean value of 177 mJy was subtracted) for two observing seasons in common with HARPS-N, taken as an example to show that the dense data sampling allows for a precise model prediction (blue curve; grey shaded area indicates the $\pm1\sigma$ confidence level) at the epochs of HARPS-N observations (indicated by red dots).}
    \label{fig:asasinterp}
\end{figure}

\begin{figure}[h]
    \centering
    \includegraphics[width=0.5\textwidth]{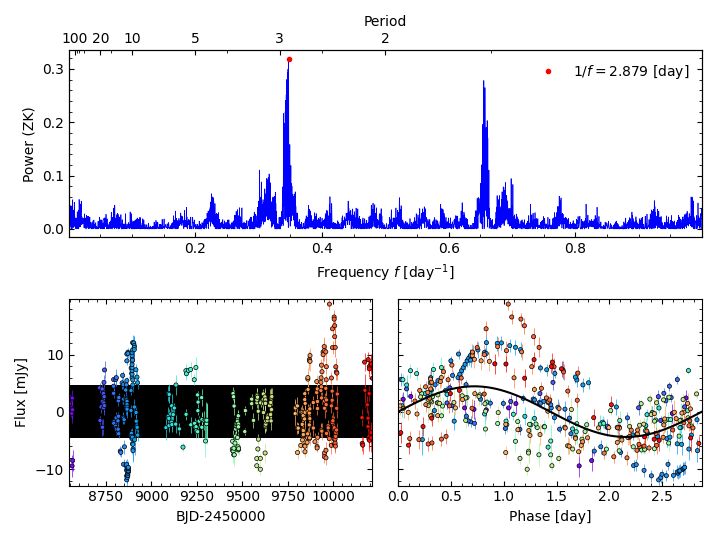}
    \caption{Periodogram and time series of the predicted ASAS-SN light curve (with the mean subtracted) at the epochs of HARPS-N observations. \textit{Top panel.} GLS periodgram. \textit{Bottom panels.} Time series (plot on the left), and light curve phase folded to the peak of the periodogram corresponding to the stellar rotation period (plot on the right). Colours are used to monitor how the phase folded light curve changes with the HARPS-N observational seasons. }
    \label{fig:glsasas2}
\end{figure}

\begin{figure*}[h]
    \centering
    \includegraphics[width=0.75\textwidth]{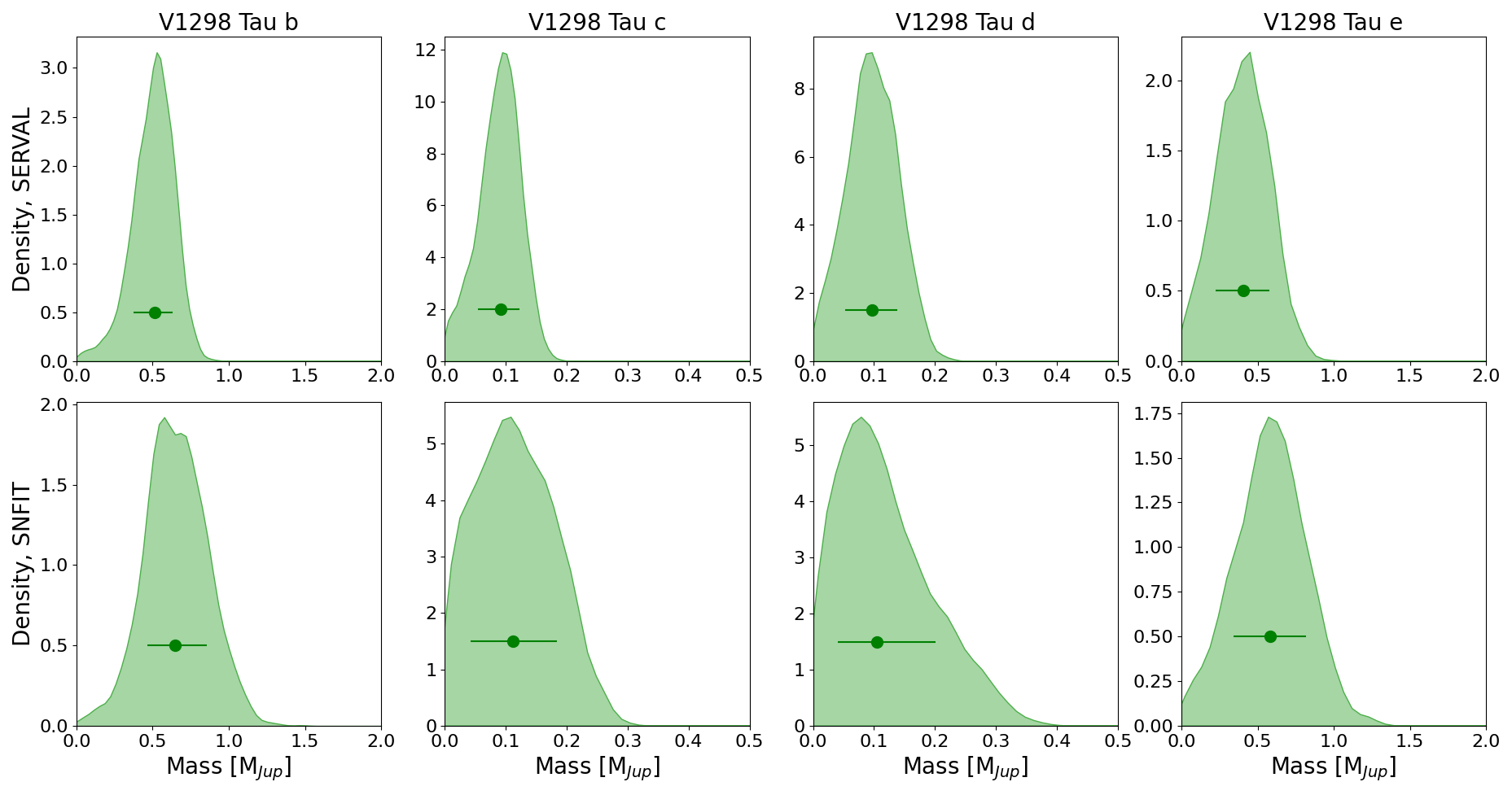}
    \caption{Posterior distributions of the masses of the four confirmed transiting planets in the V1298\,Tau, corresponding to the RV extraction methods \texttt{serval} (first row) and \texttt{snfit} (second row), and obtained through a 1D GP regression using a QP kernel to model the stellar activity contribution. The orbital motion of the planets has been modelled with the $N$-body integrator \texttt{trades}.  }
    \label{fig:masses1D}
\end{figure*}

\begin{figure*}[h]
    \centering
    \includegraphics[width=0.75\textwidth]{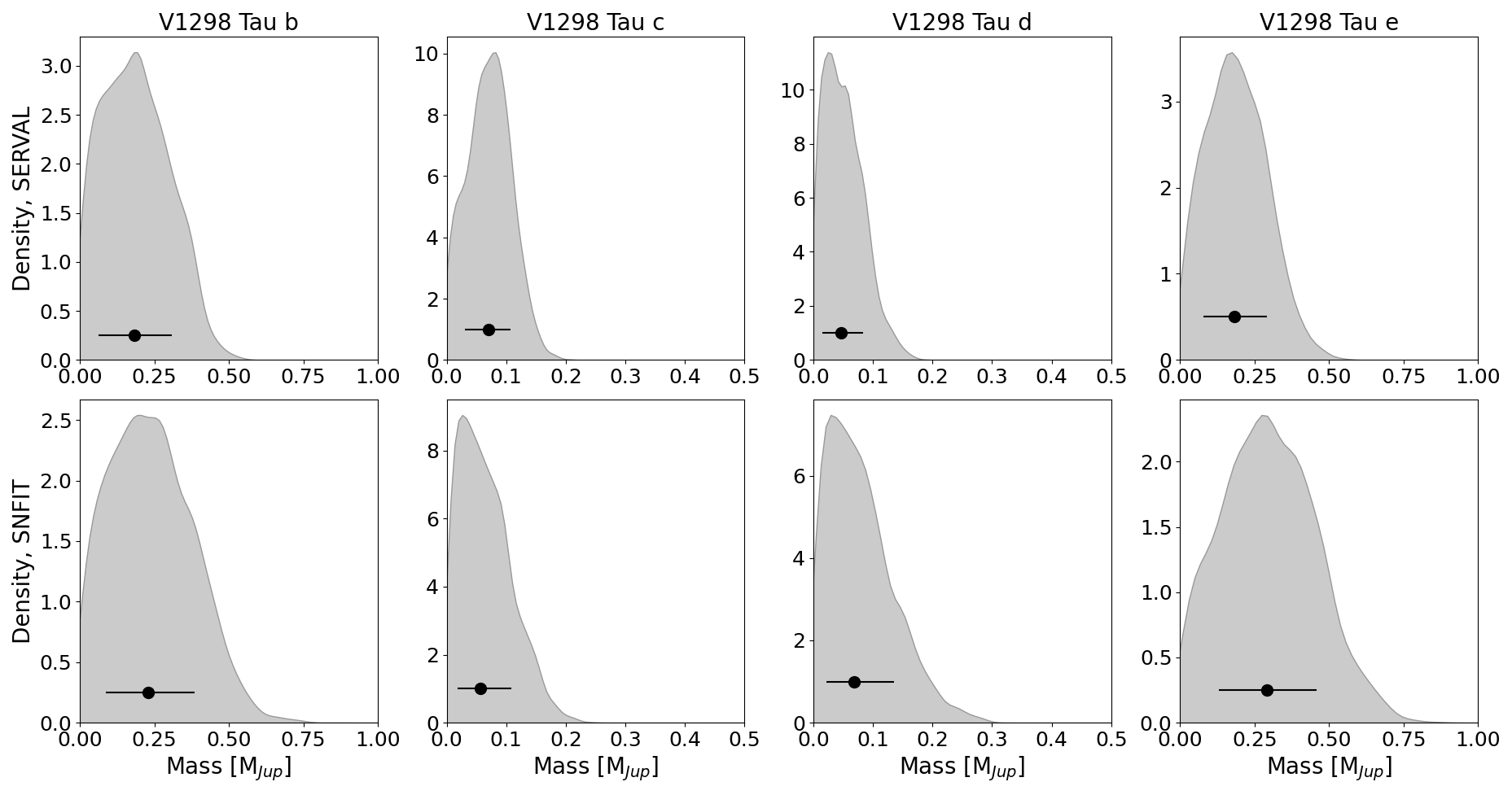}
    \caption{Same as Fig. \ref{fig:masses1D}, this time using a 2D GP regression (QP kernel) and the activity diagnostic BIS to model the stellar activity contribution.}
    \label{fig:masses2DBIS}
\end{figure*}

\begin{figure*}[h]
    \centering
    \includegraphics[width=0.8\textwidth]{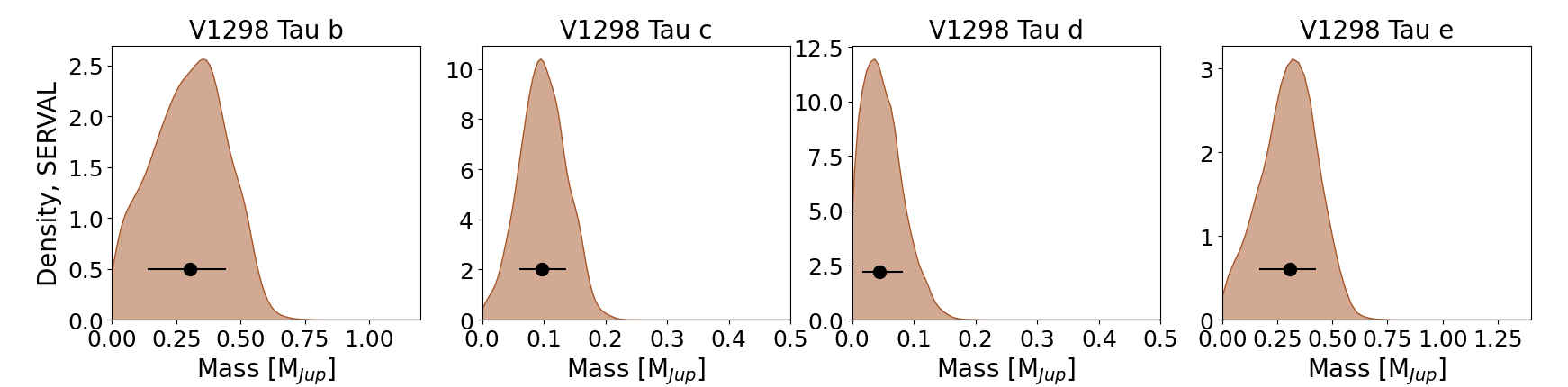}
    \caption{Same as Fig. \ref{fig:masses1D}, this time only for the RV \texttt{serval}, and a 2-dimensional GP regression (QP kernel) using the chromatic index CRX to model the stellar activity contribution. }
    \label{fig:masses2DCRX}
\end{figure*}

\begin{figure*}[h]
    \centering
    \includegraphics[width=0.8\textwidth]{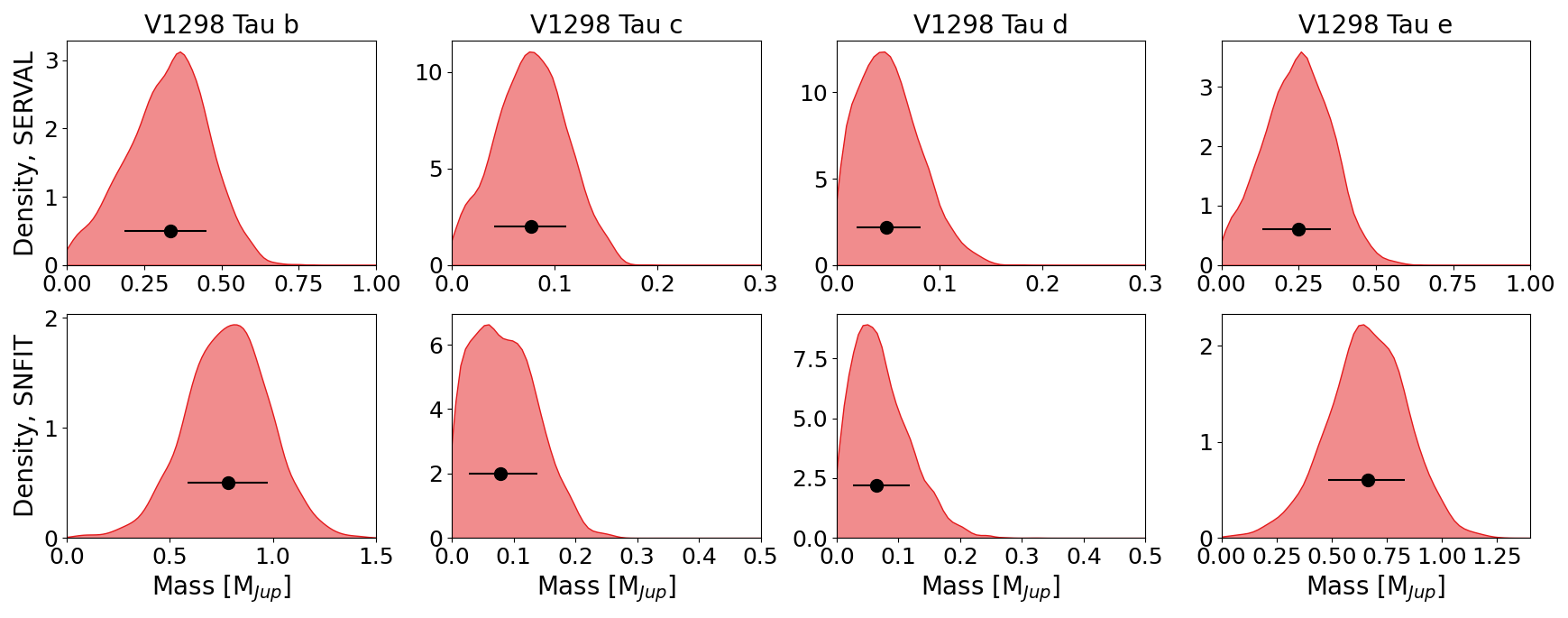}
    \caption{Same as Fig. \ref{fig:masses1D}, this time for a 1D GP regression (QP kernel) using the H$\alpha$ index.}
    \label{fig:masses2Dhalphaserval}
\end{figure*}

\begin{figure*}[h]
    \centering
    \includegraphics[width=0.75\textwidth]{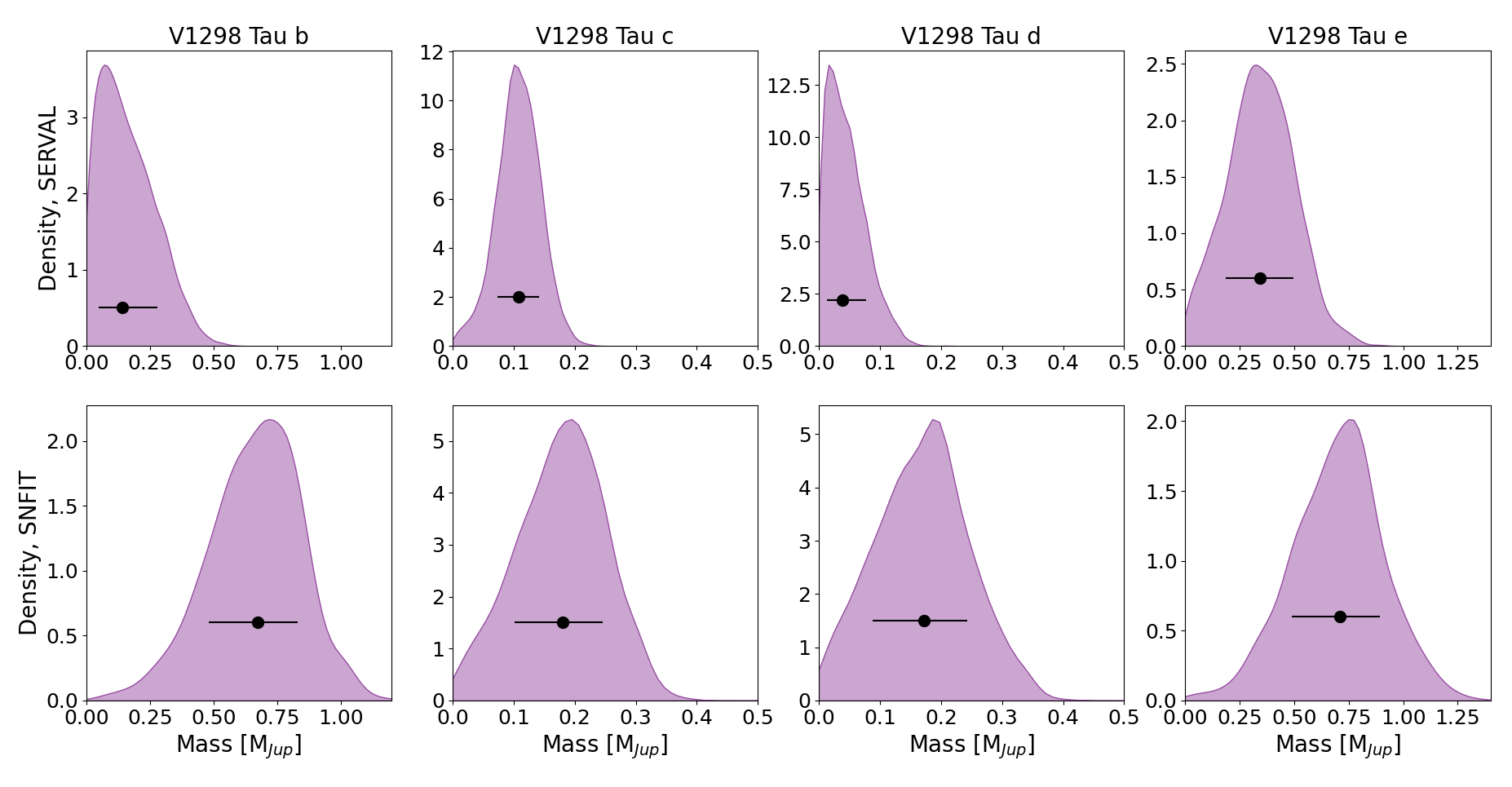}\\
    \caption{Same as Fig. \ref{fig:masses1D}, this time for a 1D GP regression (QP kernel) using the ASAS-SN photometry interpolated to the epochs of the HARPS-N observations (see Fig. \ref{fig:glsasas2}).}
    \label{fig:masses2DASAS}
\end{figure*}

\begin{figure*}[h]
    \centering
    \includegraphics[width=0.8\textwidth]{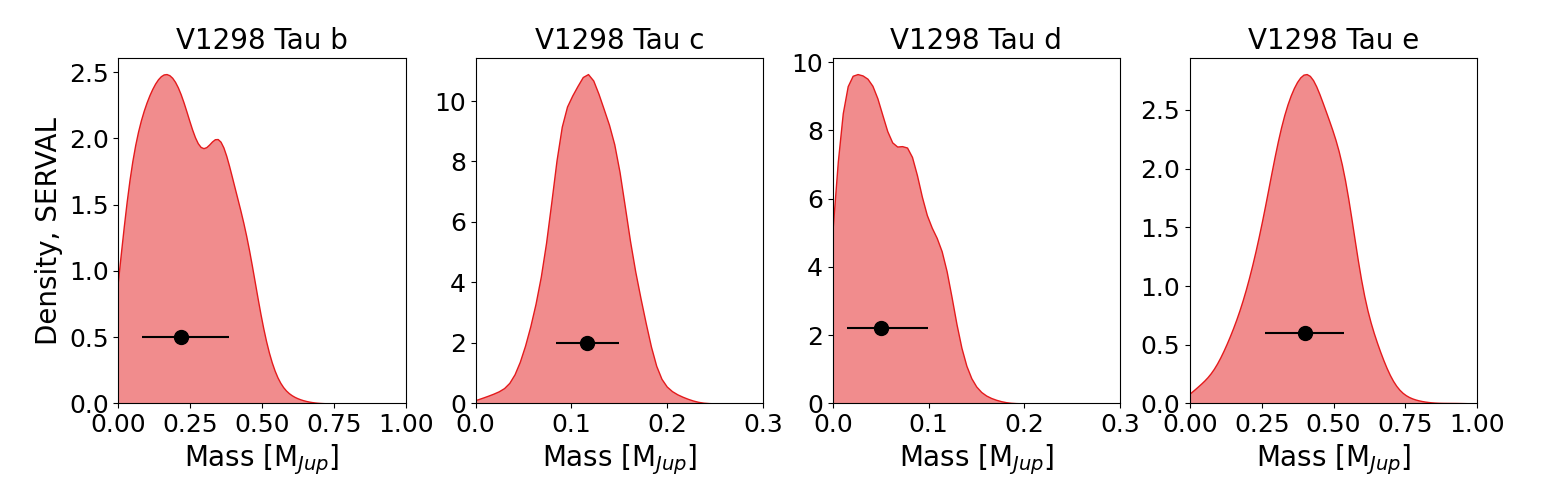}\\
    \caption{Same as Fig. \ref{fig:masses1D}, this time only for the RV \texttt{serval}, and a 3-dimensional GP regression (QP kernel), using the ASAS-SN photometry interpolated to the epochs of the HARPS-N observations, and the CRX index.}
    \label{fig:masses3DASASCRX}
\end{figure*}

\begin{figure*}
    \centering
    \includegraphics[width=0.95\linewidth]{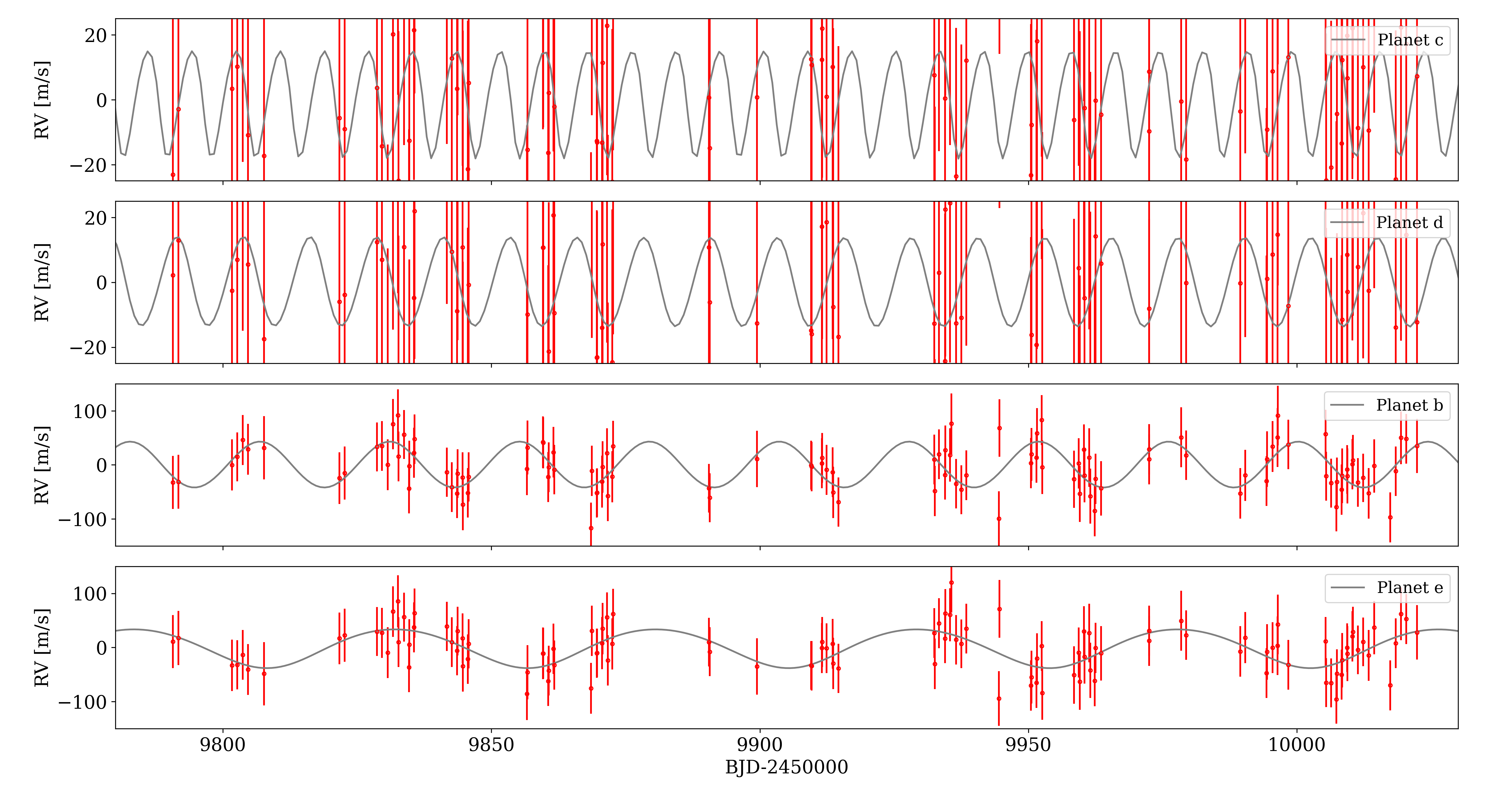}
    \caption{RV signals of the individual planets corresponding to the best-fit solution shown in Fig. \ref{fig:resultssnfitasas}. For a better readability, only the last observing season is shown.}
    \label{fig:singleplanets}
\end{figure*}

\begin{figure}
    \centering
    \includegraphics[width=0.5\textwidth]{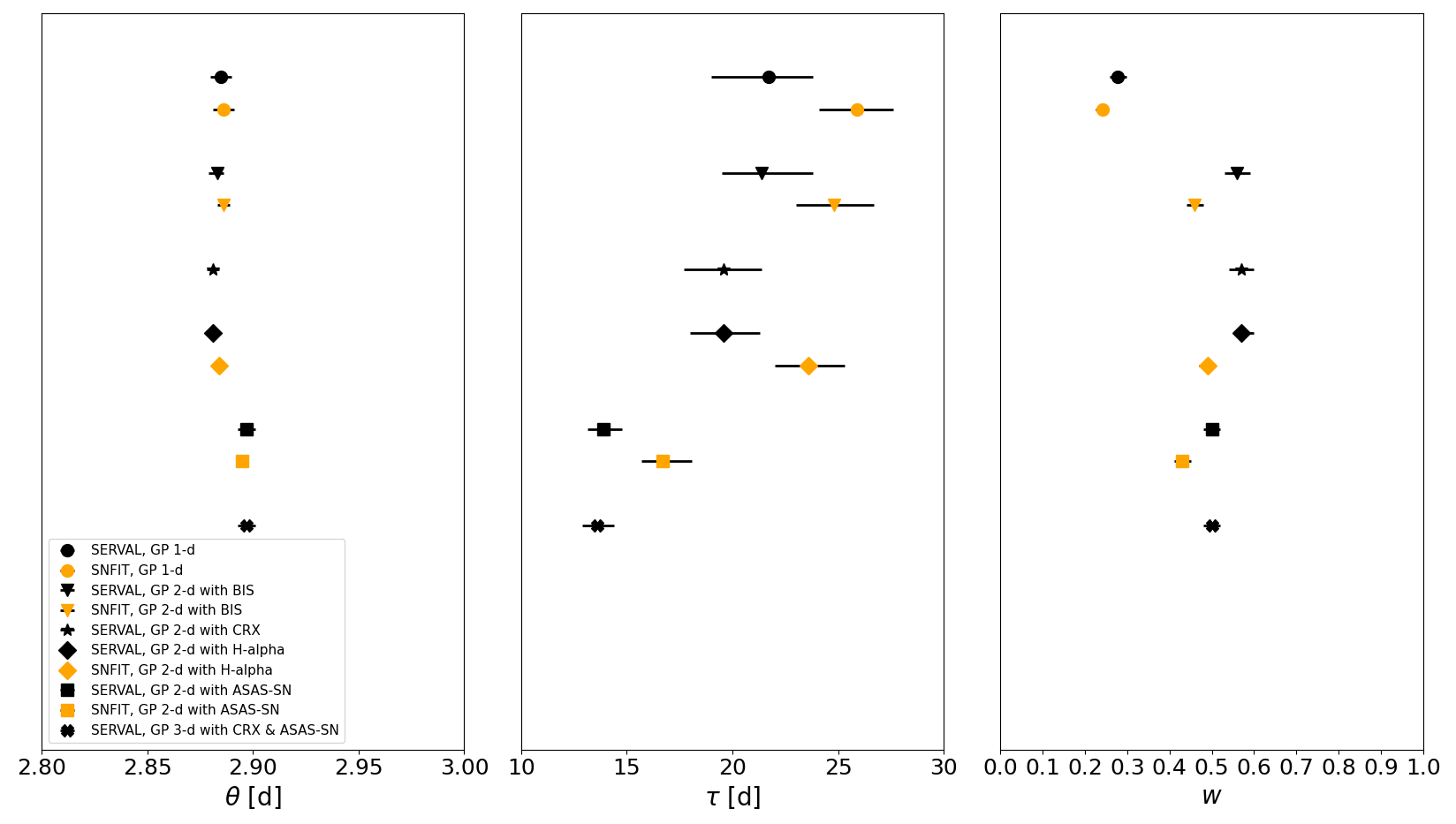}
    \caption{Best-fit values (given as the 15.87$^{\rm th}$, 50$^{\rm th}$, and 84.14$^{\rm th}$ percentiles of the posterior distributions) of the GP QP hyper-parameters $\theta$, $\tau$, and $w$ for all the test models discussed in this work.}
    \label{fig:hyperpost}
\end{figure}

\begin{table*}[!htbp]
    \centering\small\centering\renewcommand{\arraystretch}{1.1}
    \setlength{\tabcolsep}{2.0pt}
    \caption{Summary of selected published results for the four confirmed transiting planets in the V1298\,Tau system.}
    \tiny
    \begin{tabular}{l c c c c}
    \hline\hline
     Parameters & \cite{Suarez2022NatAs...6..232S} &   \cite{Finociety2023MNRAS.526.4627F} & \cite{Sikora_2023}  &  \cite{livingston2026Natur.649..310L}  \\
    \hline
    \emph{V1298\,Tau\,c} \rule{0pt}{12pt} & & & \\
    &&& \\
    Orbital period ($P$) [d] & $8.24892 \pm 0.00083$ & $8.2438_{-0.0020}^{+0.0024}$ & $8.248720 \pm 0.000024$&  $8.249164 \pm 0.000003$ \\
    Planet radius ($R_{p}$) [$R_{\oplus}$] & $5.15 \pm 0.38$ & $5.05 \pm 0.14$ & $5.24_{-0.23}^{+0.24}$ & $5.08 \pm 0.37$ \\
    Mass ($M_{\mathrm{p}}$) [$M_{\oplus}$] & $< 76$ & $< 38$ & $19.8_{-8.9}^{+9.3}$&  $4.7 \pm 0.6$ \\
    \hline    
    \emph{V1298\,Tau\,d} \rule{0pt}{12pt} & & & \\
    &&& \\
    Orbital period ($P$) [d] &$12.4058 \pm 0.0018$ & $12.3960_{-0.0020}^{+0.0019}$& $12.402140 \pm 0.000017$& $12.401394 \pm 0.000009$\\
    Planet radius ($R_{p}$) [$R_{\oplus}$] & $6.43 \pm 0.46$& $6.13 \pm 0.28$ & $6.34 \pm 0.30$& $6.53 \pm 0.42$\\
    Mass ($M_{\mathrm{p}}$) [$M_{\oplus}$] &  $< 99$ & $< 41$ &  $<36$&  $6.0 \pm 0.7$\\
    \hline
    \emph{V1298\,Tau\,b} \rule{0pt}{12pt} & & & \\
    &&& \\
    Orbital period ($P$) [d] & $24.1399 \pm 0.0015$& $24.1315_{-0.0034}^{+0.0033}$& $24.140410 \pm 0.000022$& $24.140006 \pm 0.000017$\\
    Planet radius ($R_{p}$) [$R_{\oplus}$] & $9.73 \pm 0.63$& $9.53 \pm 0.32$ & $9.95_{-0.35}^{+0.37}$&$9.41 \pm 0.57$ \\
    Mass ($M_{\mathrm{p}}$) [$M_{\oplus}$] & $203.45 \pm 60$  & $< 86$ & $<159$& $13.1 \pm 5.3$ \\
    \hline
    \emph{V1298\,Tau\,e} \rule{0pt}{12pt} & & & \\
    &&& \\
    Orbital period ($P$) [d] & $40.2 \pm 1.0$& $53.5 \pm 0.4$ & $46.768131 \pm 0.000076$&  $48.677714 \pm 0.000053$\\
    Planet radius ($R_{p}$) [$R_{\oplus}$] & $8.24 \pm 0.80 $& $9.94 \pm 0.39$& $9.50_{-0.49}^{+0.51}$& $10.17 \pm 0.75$\\
    Mass ($M_{\mathrm{p}}$) [$M_{\oplus}$] &  $369 \pm 95$ & $321_{-83}^{+111}$ & $210 \pm 82$& $15.3 \pm 4.2$\\
    \hline    
    \end{tabular}
    \label{Table:Summary_planets}
\end{table*}

\begin{table*}
    \centering
    \tiny
     \caption{Results of a GP regression analysis of RV and activity diagnostics using a QP kernel.}
    \begin{tabular}{lcccccccc}
    \hline\hline
    Parameter & Prior & \multicolumn{2}{c}{RV} & BIS & CRX & H$\alpha$ index & S$_{\rm MW}$ index & ASAS-SN\\ 
     & & \texttt{serval} & \texttt{snfit} \\
    \hline
    $h_{\rm RV}$ [\ms] & $\mathcal{U}$(0,1000) & 216$^{+13}_{-12}$ & 268$^{+18}_{-15}$ & ... & ... & ... & ... & .. \\
    $h_{\rm BIS}$ [\ms] & $\mathcal{U}$(0,1000) & ... & ... & 646$^{+33}_{-31}$ & ... & ... & ... & ... \\
    $h_{\rm CRX}$ [m s$^{-1}$ Np$^{-1}$] & $\mathcal{U}$(0,1000) & ... & ... & ... & 168$\pm$13 & ... & ... & ... \\
    $h_{\rm H_{\alpha}}$ & $\mathcal{U}$(0,5) & ... & ... & ... & ... & 0.021$\pm$0.002 & ... & ... \\
    $h_{\rm S_{\rm MW}}$  & $\mathcal{U}$(0,5) & ... & ... & ... & ... & ... & 0.074$^{+0.003}_{-0.002}$ & ... \\
    $h_{\rm ASAS-SN}$ [mJy] & $\mathcal{U}$(0,1000) & ... & ... & ... & ... & ... & ... & 4.4$\pm$0.2 \\
    \hline
    $\theta$ [d] &$\mathcal{U}$(0,10) & 2.885$\pm$0.005 & 2.885$\pm$0.003 & 2.885$\pm$0.002 & 2.877$\pm$0.003 & 3.2$^{+0.1}_{-0.3}$ & 8.81$^{+0.28}_{-0.09}$ & 2.891$\pm$0.003 \\
    $w$  & $\mathcal{U}$(0,2) & 0.29$\pm$0.02 & 0.25$\pm$0.02 & 0.154$\pm$0.009 & 0.24$\pm$0.02 & 0.38$\pm$0.09 & 0.14$\pm$0.02 & 0.28$^{+0.04}_{-0.03}$ \\
    $\tau$ [d] & $\mathcal{LU}$(0,10)& 22.0$^{+3.1}_{-2.7}$ & 25.5$^{+2.2}_{-1.9}$ & 22.8$^{+1.9}_{-1.8}$ & 38.9$\pm$6.7 & 2.5$^{+0.3}_{-0.6}$ & 9.6$\pm$2.0 & 27.0$\pm$2.0\\
     \hline 
    $\sigma_{\rm RV,\,jitt.}$ [\ms] & $\mathcal{U}$(0,50) & 23$\pm$3 & 38$\pm$4 & ... & ... & ... & ... & ... \\
    $\sigma_{\rm BIS,\,jitt.}$ [\ms] & $\mathcal{U}$(0,200) & ... & ... & 84$\pm$11 & ... & ... & ... & ... \\
    $\sigma_{\rm CRX,\,jitt.}$ [m s$^{-1}$ Np$^{-1}$] & $\mathcal{U}$(0,1000) & ... & ... & ... & 11.1$^{+11.1}_{-7.6}$ & ... & ... & ... \\
    $\sigma_{\rm H_{\alpha},\,jitt.}$,  $\sigma_{\rm S_{\rm MW},\,jitt.}$ & $\mathcal{U}$(0,1) & ... & ... & ... & ... & 0.0133$^{+0.0006}_{-0.0004}$ & 0.038$^{+0.002}_{-0.001}$ & ... \\
    $\sigma_{\rm ASAS-SN,\,jitt.}$ [mJy] & $\mathcal{U}$(0,1000) & ... & ... & ... & ... & ... & ... & 2.17$\pm$0.06\\
    \hline
    $\gamma_{\rm RV}$ [\ms] & $\mathcal{U}$(-500,500) &  -220$^{+25}_{-26}$ & -5$^{+30}_{-31}$ & ... & ... & ... & ... & ... \\
    $\gamma_{\rm BIS}$ [\ms] & $\mathcal{U}$(-2000,2000) & ... & ... & -207$^{+57}_{-61}$ & ... & ... & ... & ... \\
    $\gamma_{\rm CRX}$ [m s$^{-1}$ Np$^{-1}$] & $\mathcal{U}$(-50,50) & ... & ... & ... & 10$\pm$21 & ... & ... & ... \\
    $\gamma_{\rm H_{\alpha}}$, $\gamma_{\rm S_{\rm MW}}$ & $\mathcal{U}$(-1,1) & ... & ... & ... & ... & 0.00016$^{+0.0011}_{-0.0018}$ & 0.007$^{+0.003}_{-0.004}$ & ... \\
    $\gamma_{\rm ASAS-SN}$ [mJy] & $\mathcal{U}$(-1000,1000) & ... & ... & ... & ... & ... & ... & 177.1$\pm$0.5\\
    \hline
    \end{tabular}
    \label{tab:gpactivityindicator}
\end{table*}

\begin{table}[]
    \centering
    \caption{Results of MC simulations for the GP 1D model with the ASAS-SN photometry applied to the \texttt{snfit} RV dataset. }
    \begin{tabular}{lc}
     \hline\hline
     Parameter    &  Best-fit value$^{(a)}$ \\
     \hline
     \multicolumn{2}{c}{Planet b}\\ 
     \noalign{\smallskip }
     $P$ [d] &  24.14048$^{+0.00019}_{-0.00017}$\\
     e & 0.049$^{+0.037}_{-0.029}$ \\
     $\omega$ [deg] & 143$^{+90}_{-67}$  \\
     $\lambda$ [deg] & 46$^{+27}_{-24}$ \\
     $i$ [deg] & 88.50$^{+0.75}_{-0.82}$ \\
     $m/M_\star$ & 0.00064$^{+0.00016}_{-0.00019}$ \\
     \noalign{\smallskip }
     \hline   
     \multicolumn{2}{c}{Planet c}\\ 
     \noalign{\smallskip } 
     $P$ [d] &  8.24980$^{+0.00034}_{-0.00032}$\\
     e & 0.071$^{+0.068}_{-0.045}$ \\
     $\omega$ [deg] & 158$^{+117}_{-82}$  \\
     $\lambda$ [deg] & 298$^{+40}_{-183}$ \\
     $i$ [deg] & 89.01$^{+0.56}_{-0.70}$ \\
     $m/M_\star$ & 0.00017$^{+0.00019}_{-0.00017}$ \\
     \noalign{\smallskip }
     \hline   
     \multicolumn{2}{c}{Planet d}\\ 
     \noalign{\smallskip }
     $P$ [d] &  12.40104$\pm0.00054$\\
     e & 0.086$^{+0.055}_{-0.045}$ \\
     $\omega$ [deg] & 86$^{+148}_{-49}$  \\
     $\lambda$ [deg] & 230$^{+39}_{-58}$ \\
     $i$ [deg] & 89.04$^{+0.37}_{-0.42}$ \\
     $m/M_\star$ & 0.00016$^{+0.000066}_{-0.000076}$ \\
     \noalign{\smallskip }
     \hline   
     \multicolumn{2}{c}{Planet e}\\ 
     \noalign{\smallskip } 
     $P$ [d] &  48.68031$^{+0.00028}_{-0.00034}$\\
     e & 0.057$^{+0.060}_{-0.037}$ \\
     $\omega$ [deg] & 201$^{+95}_{-124}$  \\
     $\lambda$ [deg] & 28$^{+307}_{-18}$ \\
     $i$ [deg] & 89.04$^{+0.37}_{-0.42}$ \\
     $m/M_\star$ & 0.00066$^{+0.00016}_{-0.00020}$ \\
     \noalign{\smallskip }
     \hline   
     \multicolumn{2}{c}{GP hyper-parameters}\\ 
     \noalign{\smallskip }
     $\theta$ [d]& 2.895$\pm$0.003 \\
     $\tau$ [d] & 17.0$^{+1.4}_{-1.0}$ \\
     $w$ & 0.43$\pm0.02$ \\
     $A_{\rm RV}$, $B_{\rm RV}$ [\ms] & 25.3$^{+6.3}_{-4.8}$, $-$96.9$\pm5.4$ \\
     $A_{\rm ASAS-SN}$ [mJy] & 4.45$\pm$0.24  \\
     \noalign{\smallskip }
     \hline
     \multicolumn{2}{c}{Instrumental offsets and uncorrelated jitter}\\ 
     \smallskip 
     $\gamma_{\rm RV}$ [\ms] & 8.2$^{+5.2}_{-5.5}$  \\
     $\sigma_{\rm RV,\, jitt.}$ [\ms] & 43.7$^{+4.0}_{-4.5}$  \\
     $\gamma_{\rm ASAS-SN}$ [mJy] & 0.05$^{+0.43}_{-0.45}$  \\
     $\sigma_{\rm ASAS-SN,\, jitt.}$ [mJy] & 0.084$^{+0.082}_{-0.059}$  \\
     \hline
    \end{tabular}
    \tablefoot{
    \tablefoottext{a}{Priors are defined in Table \ref{tab:priors}. The results are calculated using the percentiles (15.87$^{\rm th}$, 50$^{\rm th}$, and 84.14$^{\rm th}$) of the posterior distributions.}}
    \label{tab:results2dasas}
\end{table}


\end{appendix}
\end{document}